\documentstyle[aaspp4,12pt,amsmath]{article}
\def \sun {$_{\scriptscriptstyle\odot}$}

\lefthead{Hungerford et al.} 

\righthead{Gamma-Rays from Single Lobe Supernova Explosions}

\received{2004 December 22}
\begin{document} 

\title{Gamma-Rays from Single Lobe Supernova Explosions} 
\author{Aimee L. Hungerford}
\affil{Transport Methods, Los Alamos National Laboratories, \\ 
  Los Alamos, NM 87545} 
\author{Chris L. Fryer\footnote{Also at:  Physics Department, University of Arizona, Tucson, AZ, 85721}, Gabriel Rockefeller$^{1}$} 
\affil{Theoretical Astrophysics, Los Alamos National Laboratories, \\ 
  Los Alamos, NM 87545} 
\authoremail{aimee@lanl.gov, fryer@lanl.gov, gaber@lanl.gov}

\begin{abstract}
  
  Multi-dimensional simulations of the neutrino-driven mechanism
  behind core-collapse supernovae have long shown that the explosions
  from this mechanism would be asymmetric.  Recently, detailed
  core-collapse simulations have shown that the explosion may be
  strongest in a single direction.  We present a suite of simulations
  modeling these ``single-lobe'' supernova explosions of a 15\,M\sun
  red supergiant star, focusing on the effect these asymmetries have
  on the gamma-ray emission and the mixing in the explosion.  We
  discuss how these asymmetries in the explosion mechanism might
  explain many of the observed ``asymmetries'' of supernovae, focusing
  on features of both supernova 1987A and the Cas A supernova
  remnant.  In particular, we show that single-lobe explosions provide
  a promising solution to the redshifted iron lines of supernova
  1987A.  We also show that the extent of mixing for explosive burning
  products depends sensitively on the angular profile of the velocity
  asymmetry and can be much more extensive than previously assumed.

\end{abstract}


\section{Introduction}
\label{sec:intro}

Observational evidence for asymmetries in core-collapse supernovae
(SNe) has been mounting for more than a decade (see Wang et al. 2001 and
Hungerford, Fryer \& Warren 2003, hereafter HFW03, for a review).
High energy observations of SN~1987A were among the first to indicate
that our assumptions of spherical symmetry were inadequate.  The
Comptonized hard X-ray continuum (Dotani et al. 1987; Sunyaev et al.
1987) was detected nearly 6 months earlier than spherically symmetric
SN models predicted, and the MeV decay lines (Tueller et al. 1990) were much
broader than what then-current theories could explain.  The earlier
emergence and broader line profiles both argued for the existence of a
small amount of radioactive nickel mixed closer to the surface of the
ejecta (see Arnett et al. 1989 for a review.)  In addition,
spherically symmetric models predicted blueshifted line centroids
(since the receding side of the SN ejecta is at large optical depth),
but the $\gamma$-ray observations showed redshifted line profiles.
Although the $\gamma$-ray data uncertainties were quite high, this
redshift was also observed in the infrared forbidden lines of [FeII]
and [CoII], providing support for the $\gamma$-ray line centroid
measurements \footnote{A caveat to keep in mind for this argument is
  that alternative mechanisms for obtaining redshifts at infrared
  wavelengths do exist.  In particular, Witteborn et al.  (1989)
  showed that electron scattering effects from a homologously
  expanding envelope (at relatively low optical depth $\sim$0.4) were
  capable of matching the redshifted lines of [ArII] (which exhibited
  similar features to the [FeII] line).}.  An early study by Grant \&
Dean (1993) demonstrated that models with lopsided $^{56}$Ni
distributions were capable of reproducing the observed $\gamma$-ray
line redshifts towards SN~1987A.  While their treatment of the
asymmetry was at the level of a toy model, and the extent of the
assumed asymmetry was rather extreme, their work demonstrated the need
for a large-scale, low mode asymmetry in the $^{56}$Ni distribution.

Observations at other wavelengths also support the existence of
large-scale asymmetry in the explosion mechanism.  Significant optical
polarization has been observed toward several core-collapse supernovae
over the past decade (Leonard \& Filippenko 2001; Wang et al. 2001),
confirming that global asymmetries are a property of the entire class
of core-collapse supernovae.  Most interesting is that these
observations show increasing polarization with time, suggesting that
it is the explosion mechanism driving these supernovae which imprints
the asymmetry (H\"oflich 1991).  Models of the nucleosynthetic yields
from SN~1987A and Cassiopeia~A supernova remnant (SNR) also
demonstrate better matches with observational element abundances when
the explosion models assume mild asymmetries (Nagataki 2000).
Furthermore, recent observations of iron and silicon X-ray line
emission (obtained with Chandra) exhibit clear morphological evidence
of global asymmetries in the distributions of these elements within
the Cas~A SNR (Hwang et al 2004).

Finally, an equally compelling argument for global asymmetry arises
from attempts to understand the high space velocities of neutron
stars.  The high observed velocities of pulsars and the formation
scenarios of neutron star binaries both suggest that neutron stars are
given strong kicks at birth.  These kicks are most easily explained by
some asymmetry in the supernova explosion (see Fryer, Burrows, \& Benz
1998 for a review).  Observations of the pulsar shaping the guitar
nebula place its proper motion measurement in excess of
$\sim$1000~km/s for the neutron star (Chatterjee \& Cordes 2002).
While its velocity places this neutron star near the high end of the
observed pulsar velocity distribution, it is not off the charts.  The
distribution of pulsar birth velocities is bimodal, with peaks at
90~km/s and 500~km/s, where each peak represents roughly 1/2 the
total population (Arzoumanian et al. 2002).

While the neutron star velocities and the spectral data both argue for
the presence of a global explosion asymmetry, theoretical studies tell
us that the origin of the asymmetries may not be related.  Many kick
mechanisms do not necessarily require asymmetries in the explosion
itself (e.g. the neutrino driven mechanisms of Arras \& Lai 1999;
Fryer 2004).  Conversely, the bipolar explosions produced by
Fryer \& Heger (2000) can explain some of the observed extended mixing
in SN~1987A (HFW03) and may even explain the asymmetric appearance of
Cas~A (Hwang et al. 2004).  However, they can not explain high neutron star
velocities.  

Scenarios capable (in principle) of generating kicks and ejecta
asymmetries simultaneously do exist, though.  In fact, the first
multi-dimensional calculations of the collapse of a massive star to an
explosion motivated just such a mechanism.  Herant et al. (1994)
argued for the importance of convection in enhancing the conversion of
neutrino energies into kinetic explosion energy.  One year later,
Herant (1995) argued that if the convective cells could merge into one
uprising bubble and one downflow, the explosion ejecta would be
sufficiently asymmetric to explain neutron star kicks.  Recent
instability analyses by Blondin et al. (2003) and full core-collapse
calculations by Scheck et al. (2004) have shown that such low mode
convection may well be possible.  In particular, Scheck et al. (2004)
found that asymmetries arise naturally within the standard model of
convective, neutrino-driven supernovae if the explosion is
sufficiently delayed.  They find uni-polar or single-lobe asymmetries
capable of imparting kicks from 0~km/s to 1500~km/s.  

The range of explosion asymmetries which result in neutron star kicks
of this magnitude appear relatively mild (i.e. the explosion energies
along the single-lobe are a few times larger than those in the
remaining parts of the star.)  This is in contrast to the explosion
asymmetries in the magnetic, jet-driven supernovae proposed by
Khokhlov et al. (1999), where all of the energy in the explosion is
injected along a bipolar jet-axis.  Wang et al. (2001) have argued
that such mild asymmetries are unlikely to produce the global density
asymmetries needed to generate the observed polarization.
Parameterized 2D models of core explosions (carried out by Chevalier
\& Soker 1989) also show a tendency for explosion asymmetries to
spherize as the shock travels through the shallow density gradient in
the hydrogen envelope of a SN Type~II progenitor.  Since polarization
is observed to be a general property of core-collapse supernovae (both
in Type~II and Type~Ib/c events), it is important that global density
asymmetries persist through shock breakout even in progenitors with
massive hydrogen envelopes (which stellar evolution models show to be
the most common SN progenitor type.)

In this paper, we address both the issue of lopsided ejecta, and the
persistence of global density aysmmetries in the explosion of a {\it
  {standard} } supernova progenitor.  This provides a baseline study
upon which future work focussing on specific progenitors (e.g. as
determined from observations of SN~1987A, Cas~A) can build.  We extend
the work of HFW03 by concentrating on single-lobe explosion
asymmetries and their effects on supernova observations.  Our
implementation of these single-lobe explosions is guided by 1) the
magnitude of the velocity asymmetry and 2) the angular profile of the
velocity asymmetry obtained from realistic models of core-collapse
supernovae (Fryer \& Heger 2000; Scheck et al. 2004).  Specifically,
we show that redshifted $\gamma$-ray lines can be produced entirely
with the low-mode velocity asymmetries currently present in numerical
core-collapse models (Scheck et al.  2004).  Furthermore, the steep
angular profile in velocity for these asymmetries allows global
density asymmetries to persist even after the shock passes through the
envelope of our red supergiant (RSG) progenitor model.  This results
in ejecta morphologies that are reminiscent of those observed in the
Cas~A SNR.  Since we do not study the specific progenitors of SN~1987A
or SNR Cas~A, we can not say definitively that the current asymmetries
in core-collapse calculations will explain the observations of these
objects.  However, these simulations do provide ejecta structures
which could be used as the building blocks to construct the specific
features seen in SN~1987A and SNR Cas~A observations.

In \S 2, we present our asymmetric explosion calculations along with
the numerical schemes and initial conditions used for those
calculations.  In \S 3, we discuss the effects such asymmetries have
on the nuclear yields from these supernovae.  Our calculations of the
high energy (X-ray and $\gamma$-ray) emission and the resulting
observations are presented in \S 4.  We conclude with a discussion of
the application of these calculations to supernova asymmetries
observed in SN~1987A and SNR Cas~A (including a discussion listing
future work that will truly determine the efficacy of convection
asymmetries in explaining the observed asymmetries in these objects).

\section{Explosion Simulations}

For our parametric study of single-lobe explosion asymmetries, we use
the same 15\,M\sun\ progenitor star (s15s7b2 from Weaver \& Woosley
1993) used in HFW03.  We also model the explosions using the same
incarnation of the 3-dimensional SNSPH code (Fryer \& Warren 2002)
used in HFW03 with only a few modifications described below.  Low mode
asymmetries in the explosion will lead to lopsided nickel
distributions and may also alter the synthesis of nickel due to the
angular variation in explosion energy.  Because we start our
simulations 100\,s after the launch of the explosion, no further
nuclear burning occurs, and we can only estimate the possible effects
asymmetric explosions might have on the actual synthesis of the ejecta
elements.  Instead, the simulations presented here concentrate on the
hydrodynamic mixing of these elements.

\subsection{Numerical Schemes}

\begin{figure}[htp]
\plotone{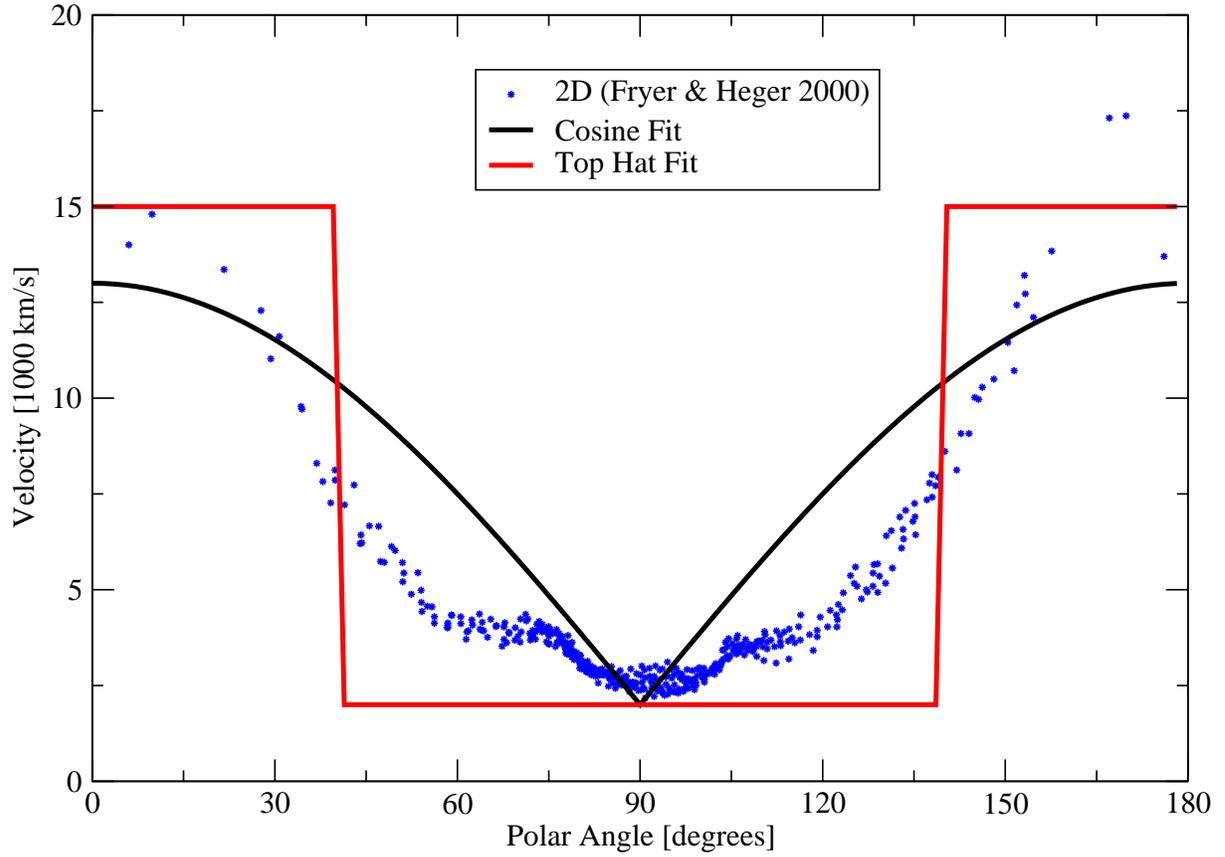}
\caption{Plot of radial velocity versus polar angle for the 2D
rotating collapse model of Fryer \& Heger (2000).  Overplotted are the
cosine function used in HFW03 and the top-hat function to represent
the profiles for the artificial velocity asymmetries.  Note that the
model profile lies somewhere between these two extremes. }
\label{fig:2dvelvsphi}
\end{figure}

Roughly 1.2 million variably massed particles are used in the SPH
simulation.  Instead of using a grid of smooth particle hydrodynamics
particles (HFW03), the neutron star (with mass of 1.4\,M\sun) is cut
out of the simulation.  Its gravitational effect is mimicked with a
central gravitational force term\footnote{This is not a
self-consistent treatment as the neutron star will have some velocity
associated with it in order to conserve momentum in these one-sided
explosions.  The ejecta which are important for our $\gamma$-ray
studies are homologously expanding with velocities greater than the
escape velocity within a few weeks after the explosion is launched.
The fastest moving neutron star in our simulations has a position
offset from the center of the ejecta of roughly 20\% the radius.  This
does not significantly affect the outflow trajectory of the ejecta
that has already received escape velocity.  In addition, this offset
corresponds to material not yet probed by the escaping $\gamma$-ray
emission, and can be neglected for this epoch (t=365~d).}.  This reduces the
numerical noise in the core and relaxes the Courant constraint on the
timestep.  As with the bipolar asymmetries, the single lobe asymmetry
is imparted by artificially altering the explosion velocities at
100\,s after bounce.  These input asymmetries require a factor by which the
velocity is enhanced and an angular profile describing the assumed
structure of the asymmetry.  In HFW03, the angular dependence of the
imposed velocity asymmetry was chosen to be sinusoidal.  Their
decision was made primarily to facilitate closer comparison with the
results of Nagataki (2000) who also assumed this smoothly
varying cosine behavior.  Comparing with the 2D rotating collapse
simulations of Fryer \& Heger (2000) and the single lobe simulations
of Scheck et al. (2004), we find that a sinusoidal velocity profile is
probably smoother than the actual profiles in the hydrodynamic models.
Figure~\ref{fig:2dvelvsphi} shows the SPH particle velocities from
Fryer \& Heger (2000) with a sinusoidal fit overplotted.  A top-hat,
with its sharp transition, represents the other extreme in fitting the
asymmetries from the explosion models (Figure~\ref{fig:2dvelvsphi}).  As we
will discuss in the next section, a flat velocity profile in angle 
along the enhanced explosion axis significantly affects the outward mixing
of nickel.  While the steep transition of the top-hat profile may be 
more extreme than the numerical models, the existence of a flat-topped velocity
profile (over polar angles of roughly 20 degrees) is likely more realistic
than the sinusoidal profiles given the current core-collapse model results
(Scheck et al. 2004; Fryer \& Heger 2000).

Our set of single lobe explosion asymmetries are created assuming the
discontinuous top-hat distribution, which represents a conical
geometry and can be described by two primary parameters: 1) {\bf
  $\Theta$} ~=~opening angle of the enhanced explosion cone, and 2)
{\bf \it f} ~=~ratio of in-cone velocity to the corresponding
out-of-cone velocity (the model names are meant to reflect these
parameters; e.g. model f2th20 has {\bf \it f}~=~2, {\bf
  $\Theta$}~=~20).  The in-cone and out-of-cone radial velocities are
determined by keeping the same kinetic energy as the symmetric
explosion while forcing {\bf \it f} to a chosen value.  They are given
by:

\begin{equation}
V_{in-cone} = {\rm \it f}\left[{\frac{1-f^2}{2}}cos(\Theta) +
{\frac{1+f^2}{2}}\right]^{-\frac{1}{2}}~V_{symm}
\end{equation}

\noindent and

\begin{equation}
V_{out-of-cone} = \left[{\frac{1-f^2}{2}}cos(\Theta) +
{\frac{1+f^2}{2}}\right]^{-\frac{1}{2}}~V_{symm}
\end{equation}

\noindent where $V_{symm}$ is the radial velocity from the 
input 1D calculation.

Table 1 lists the suite of simulations studied in this paper.  We have
focused on an opening angle of 20$^\circ$, but include a symmetric
model with identical neutron star set-up for comparison.  Additionally,
a {\bf $\Theta$}=40$^\circ$ model was run to get a rough impression
of the dependence on opening angle.  We have also included two
simulations where the $^{56}$Ni abundance has been enhanced by 100\%
within the opening angle.  This corresponds to the maximum increase in
the $^{56}$Ni abundance seen in super-energetic explosion calculations
(Hungerford et al. 2004).  Because the opening angles are small, this
enhancement corresponds to an increase in the total $^{56}$Ni abundance 
of less than $\sim$10\%.

As mentioned above, our choice of parameter values for this study are
guided by the multi-dimensional simulations of Scheck et al. (2004)
and Blondin et al. (2003), which find low mode instabilities driving
the SN explosion.  These calculations have found that low-mode
instabilities can produce kick velocities ranging from 100~${\rm km
  s^{-1}}$ and 1700~${\rm km s^{-1}}$.  Although our simulations also
assume a range of kick velocities from 0-1690~${\rm km s^{-1}}$, our 
simplified models (both the single-lobe explosion and top-hat
transition function) do not model the full range of results produced
by the Scheck et al. (2004) and Blondin et al. (2003) calculations.
The simulations in this paper are meant to give a flavor of what
we can expect from asymmetries.  A more thorough study of the range of
asymmetric effects on the mixing will be presented in a later paper.

\subsection{Explosion and Nickel Distribution}
\label{sec:expni}

\begin{figure}[htp]
\plotone{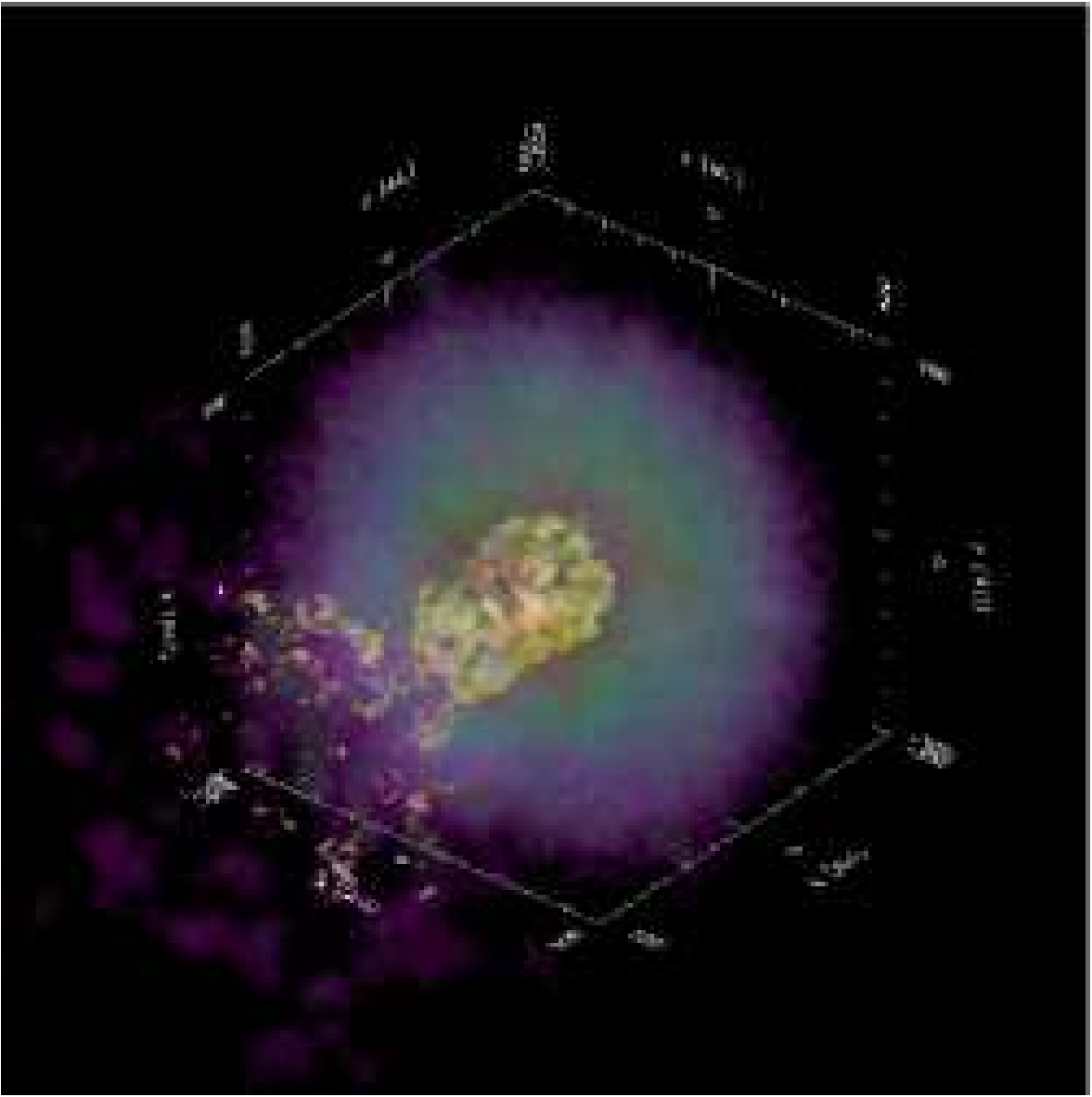}
\caption{3-dimensional rendering of the f3th40 explosion model 1 year
after the shock launch.  The isosurface represents the cobalt
distribution with a number density of $10^{-5}$.  The colors denote
the density distribution.  The top-hat distribution of the imposed 
velocity asymmetry allows a significant portion of the cone material
to expand without drag from fluid shear forces, resulting in the
large splash of material at the outer ejecta.
}
\label{fig:f3th403D}
\end{figure}

\begin{figure}[htp]
\plotone{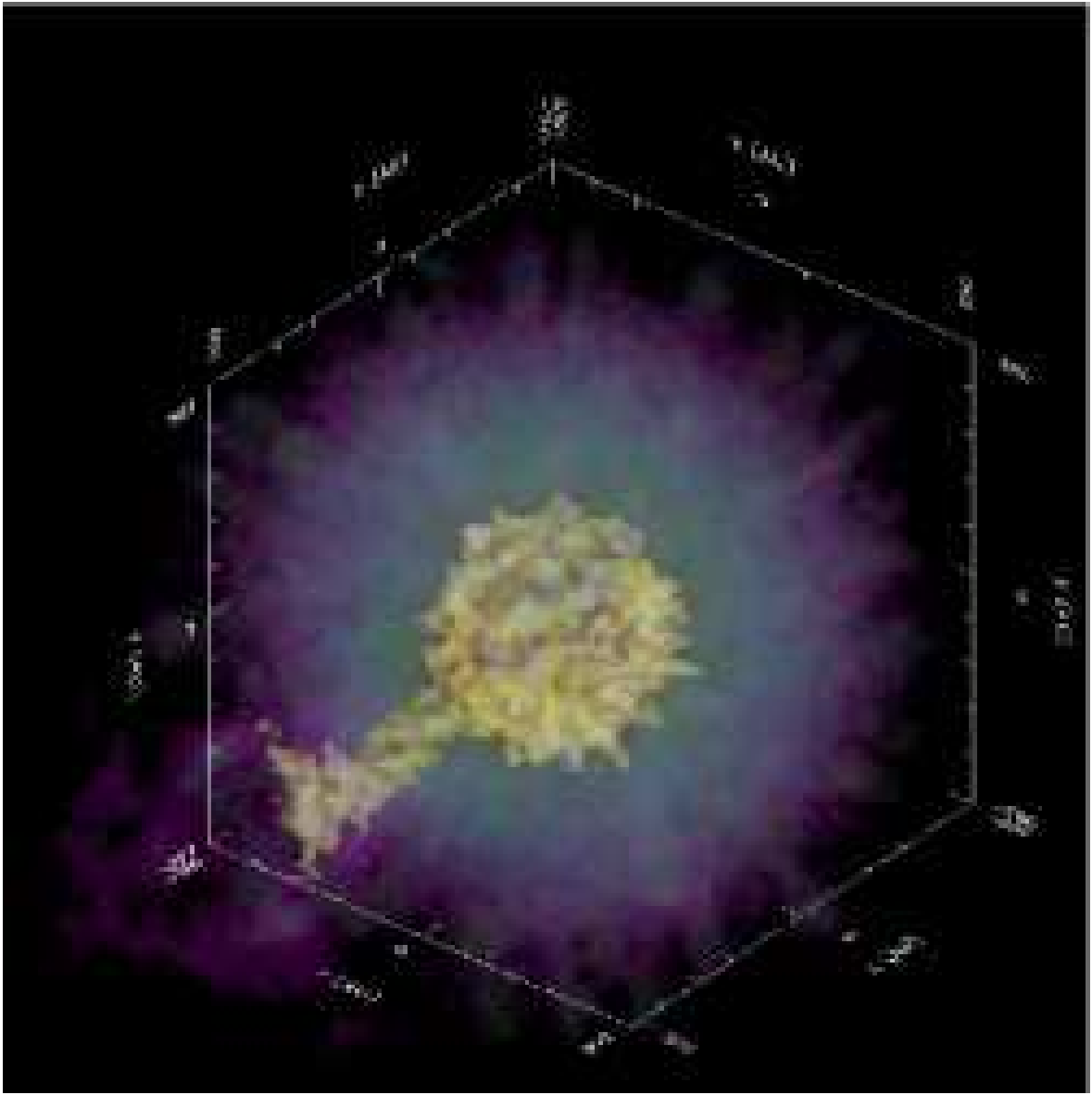}
\caption{Same as previous figure, but for model f2th20.  Again we
see the splash of cobalt in the outer ejecta, but it is not as extreme
because of the smaller angle and lower contrast of the velocity asymmetry.}
\label{fig:f2th203D}
\end{figure}

Figures \ref{fig:f3th403D} and \ref{fig:f2th203D} show the $^{56}$Co
distribution (isosurfaces of number density set to $10^{-5}$) within
the density distribution of the exploding star (shading) for the
f3th40 and f2th20 models.  These figures show the global view of the
ejecta morphologies which develop from two extremes in our set of parameterized
explosions.  In particular, it is clear that the density asymmetries
(shading) in both of the single lobe explosion models are much more
extreme in these simulations than for the bipolar explosions of HFW03 
(see their Fig. 5).  The 3:1
velocity asymmetry of model f3th40 is sufficient to push through the
entire star, creating a spray of heavy elements at the outermost edge
of the expanding ejecta.  Even the 2:1 velocity asymmetry in model
f2th20 leads to aggressive outward mixing of nickel and its decay
products, placing them very near the edge of the stellar ejecta as
well.

The peak magnitudes of these velocity asymmetries do not differ
significantly from those of the bipolar explosions in HFW03, but the
outward mixing of the $^{56}$Co is much more extensive than HFW03
found in their bipolar explosions.  This is primarily due to the
difference between the angular profiles of the velocity asymmetry.  As
mentioned above, the bipolar explosions were given a smoothly varying
cosine velocity asymmetry, while the single lobe explosions assume a
discontinuous ``top-hat'' profile in polar angle.  The flat angular
profile in the velocity for these single lobe models allows a
significant portion of the high velocity region to expand without drag
from fluid shear forces.  This likely the cause of the increased
heavy element penetration in the single lobe explosion models, rather
than any differences in the geometry of the asymmetry.  That said, results
from current core-collapse models show that such flat-topped profiles
in angle do exist over polar angles of 20 to 40 degrees 
(Figure \ref{fig:2dvelvsphi}; also see Figure 2 in Scheck et al. 2004).

\begin{figure}[htp]
\plotone{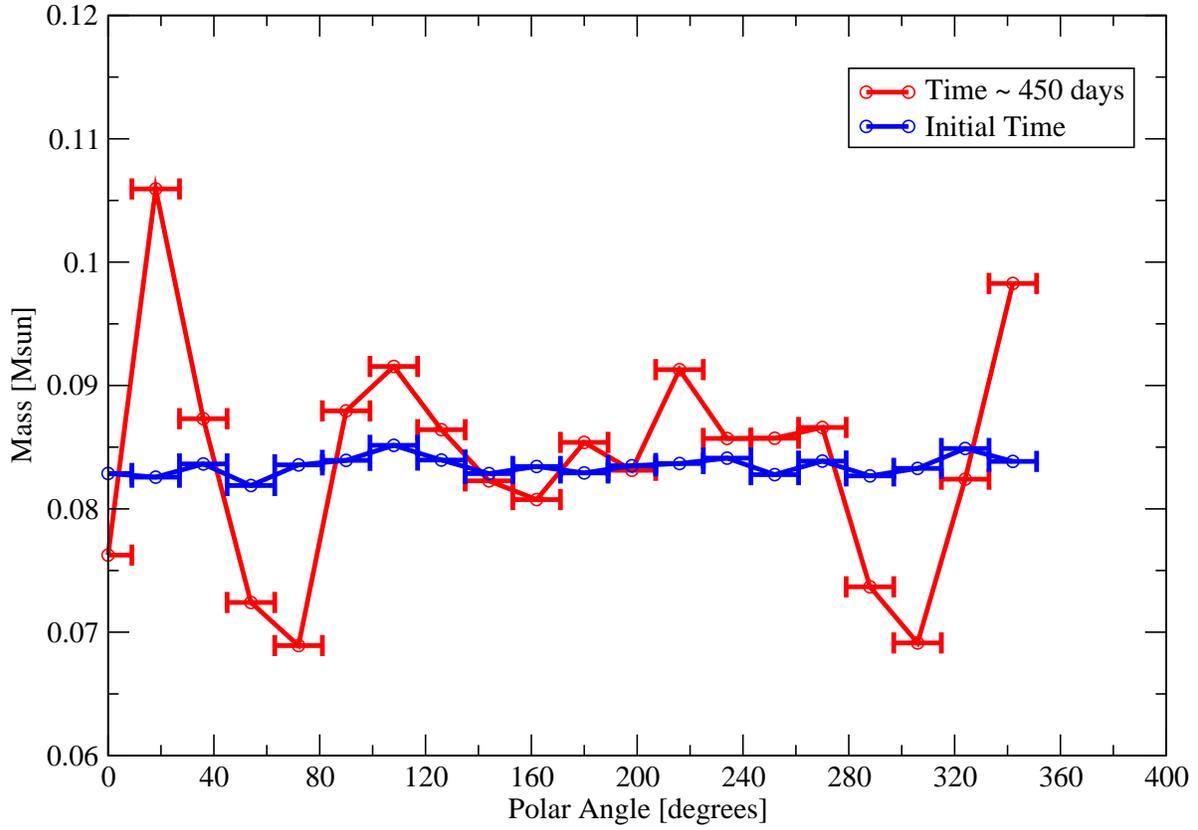}
\caption{Plot of mass in a cone of radius 9$^\circ$ along a direction in
polar angle for the f3th40 model.  Error bars reflect the cone diameter.  
Blue line is for the initial time at t~=~100~s and red line is for 
t~=~450~days (long after flow has become homologous.)  Matter is 
being funneled into the faster expanding, lower density region in the
cone.  This results in an enhancement of the mass for polar angles near zero,
and a reduction in mass for angles just outside of this.}
\label{fig:mass_f3th40}
\end{figure}

A more quantitative look at the mass motions and energetics of the
explosions can be seen in
Figures~\ref{fig:mass_f3th40}~-~\ref{fig:energy_f2th20} where the
kinetic energy and mass are shown in cones with a 9$^\circ$ angular
radius for a variety of polar angles in the f2th20 and f3th40
explosion models.  The blue curve in each plot denotes the initial
conditions of the model (100~s after launch of explosion), and the red
curve is at later times when the ejecta flow has become homologous.
From the mass distributions, we see that the mass in the enhanced
explosion lobe has increased while the mass at angles just outside the
cone has been reduced.  The in-cone material is shocked to higher velocities
and expands rapidly, creating at its back a low density wake into which the
surrounding material is funneled, thus increasing the total mass in
the cone region at the expense of the material just outside the cone.
This migration of material to the high velocity cone region affects
the heavy element distribution as well, resulting in a larger mass of
radioactive nickel within the cone.  For the specific case of model
f3th40, the enhanced explosion lobe is sufficiently energetic to poke
a hole out of the star, spraying the matter from this cone in all
directions.  This leaves it with an overall enhancement in mass over
the enhanced velocity cone, but actually results in a reduction of
material right along the cone axis.  The f2th20 model was not
energetic enough to poke through the edge of the envelope, so it
retains the mass enhancement over the entire cone region.

In conjunction with the mass distributions, the energy distributions
show that a spherization of the explosion is taking place.  While the
mass in the cone has increased over time, the kinetic energy of the
cone material has actually dropped off.  This lost energy is spreading
to the rest of the ejecta, equalizing the explosion velocities in
angle.  In the bipolar explosions of HFW03, the spherization of the
explosion was essentially complete once the shock had passed through
the star.  For the explosions presented here, the peak velocity along
the pole remains larger than the equatorial velocity, even after the
shock has passed through the entire star.  This velocity asymmetry
gives rise to asymmetric density contours as shown in
Figures~\ref{fig:f3th40_nidist}~and~\ref{fig:f2th20_nidist} (blue
contours in the left center panel) for models f3th40 and f2th20,
respectively.  Such persistent density asymmetries may be necessary to
explain observations of supernova remnants.  For instance, the
morphology of SNR Cas A (with its optical jet) requires a density
asymmetry that survives not just propagation through the star, but
even significant interaction with the interstellar
medium\footnote{Keep in mind that it is also possible that the remnant
  asymmetries are caused by interaction with the circumstellar medium
  and have little to do with the explosion mechanism itself.}.

While our chosen velocity asymmetries seem to reproduce features from
remnant morphology, high energy line and continuum observations from
SN~1987A possessed trends (broad lines, early detection of continuum)
that must also be reproduced by these extended mixing structures.  A
first order test of the adequacy of this mixing is to compare our
theoretical nickel distributions, plotted against line-of-sight
velocity, with features in the observed infrared line profiles.  Haas
et al. (1990) argued that the [FeII] line profile reflects the
entirety of the spatially distributed radioactive nickel, as any
optically thick component of iron is likely distributed throughout the
ejecta in clumps.  The panels in
Figures~\ref{fig:f3th40_nidist}~and~\ref{fig:f2th20_nidist} show
nickel distributions versus line-of-sight velocity for models f3th40
and f2th20 at a variety of viewing angles as indicated by the black
direction vectors.  In addition, the distributions for the set of
fXth20 models are plotted in Figure~\ref{fig:nidist} for the
anti-polar and equatorial viewing angles.  For all these plots, the
broadest line profiles are obtained when looking along the axis of the
explosion asymmetry, though the range in profile shapes is quite
large.  Models f5th20 and f3th20 possess redward wings which do result
in broadened line profiles (relative to the Symmetric model) as
suggested by SN~1987A data.  The f3th40 and f2th20 show relatively
narrow line cores, however, they both possess a bump containing a few
percent of the total nickel mass at high velocity.  This high velocity
component is particularly interesting as such a ``bump'' was invoked
to explain irregularities in the H-$\alpha$ line observations of
SN~1987A (i.e.  ``the Bochum event'').

The extended mixing in these simulations is also reflected in the
spatial distribution of individual elements.  Figure~\ref{fig:encmass}
shows plots of $^{56}$Ni (top panel) and $^{44}$Ti (bottom panel)
versus enclosed mass for our suite of explosion models.  Note that
3-dimensional mixing in our symmetric explosion places some nickel
right up to the helium layer, well beyond the inner layers of the
ejecta.  In the asymmetric simulations, this mixing is much more
extreme, placing nickel out to the outer layers of the star with a
hundredth of a solar mass or more of nickel mixed into the hydrogen
layer!  If we assume that the inner 3-5\,M\sun falls back to form a
black hole, we still eject $\sim$0.01-0.03\,M\sun nickel to power a
weak supernova explosion\footnote{Bear in mind that in our simulation,
starting 100\,s after bounce, there is very little fallback.  To really
determine the mixing for an explosion with this much fallback, we would
need to model an explosion with weaker energies starting at earlier times.
Plans for such calculations are underway.}.  If this much fallback occured
in our symmetric explosions, no nickel would be ejected and the optical
display from this explosion would be too weak to be observed.  Optical
displays from systems that we believe form black holes are a sure sign
of supernova asymmetry.

\section{Abundance Enhancements}

In \S \ref{sec:expni}, we showed how asymmetries can mix out the
nickel formed in the explosion, transporting some of this nickel from
the depths of the star well into the hydrogen envelope.  Asymmetries
in the explosion energy 
can also alter the abundance of elements synthesized in the explosion.
Because we do not start our simulations until 100\,s after the launch
of the explosion, we can not directly calculate the yields from these
explosion asymmetries.  But we can estimate the effect of asymmetries
on the element synthesis by combining the results from 2 separate
1-dimensional simulations to mimic both the strong conical burst
and the remainder of the star.  For this estimate, we use the
1-dimensional explosions of a 15\,M$_\odot$ star from (Hungerford et al.
2004, 2005).  In the most extreme case (our f5th20 model), the energy
within the lobe corresponds to a spherical explosion of $\sim 15 \times
10^{51}$\,ergs (although its 1\% total volume means that it
contributes only $1.5 \times 10^{50}$\,ergs to the total explosion
energy of this asymmetric outburst).  The rest of the star has an
effective spherical energy of roughly $0.6\times 10^{51}$\,ergs.

For the f3th40 simulation, the energy within the lobe corresponds to
$\sim 5 \times 10^{51}$\,ergs, but the total volume is 10\% and it is
this model that will show the largest variation in nucleosynthetic
yield from supernova asymmetries.  Figure \ref{fig:abund} (top) shows
the abundance fraction (relative to solar) of stable isotopes as a
function of isotope mass for 1-dimensional explosions with energies
of $1.35 \times 10^{51}$\,ergs (black) and $6.5 \times 10^{51}$\,ergs
(blue).  The red data shows the linear combination (90\% at $1.35
\times 10^{51}$\,ergs, 10\% at $6.5 \times 10^{51}$\,ergs).  The
abundances are only slightly different than what we would expect from
a 1-dimensional explosion with an energy of $1.35 \times
10^{51}$\,ergs.

For the strong explosions in this paper, the effect on the
nucleosynthetic yields of stable isotopes is minimal.  But the
situation is very different for some radioactive yields.  Let's 
look at these yields in more detail.  
\begin{itemize}
\item{\bf $^{56}$Ni}: $0.014,0.154, 0.37$M$_\odot$ for $0.1,1.35,6.5
  \times 10^{51}$\,erg explosions.
\item{\bf $^{44}$Ti}: $<10^{-20},1.3 \times 10^{-5}, 1.4 \times
  10^{-6}$M$_\odot$ for $0.1,1.35,6.5 \times 10^{51}$\,erg explosions.
\item{\bf $^{26}$Al}: $<10^{-20},3.4 \times 10^{-10}, 4.3 \times
  10^{-5}$M$_\odot$ for $0.1,1.35,6.5 \times 10^{51}$\,erg explosions.
\end{itemize}
Especially notice the yield of radioactive $^{26}$Al.  The $^{26}$Al
yields reported above do not include the hydrostatically synthesized
contribution, however recent models from Limongi \& Chieffi (2005) argue that
explosive production of $^{26}$Al dominates the total yield for a wide
range of main sequence masses; this is particularly true at the lower
masses like the 15 solar mass progenitor employed here.  Our results
show that the $^{26}$Al synthesized during the explosion can be orders
of magnitude higher in an asymmetric explosion than in a symmetric
explosion of comparable total explosion energy.  Depending upon the
frequency and magnitude of explosion asymmetries, this enhancement may
allow SN~II to be the primary synthesis site for $^{26}$Al, a
possibility which had been ruled out based on theoretical yields
assuming spherical symmetry (see Higdon, Lingenfelter \& Rothschild
2004 and references therein).  However, be aware that these estimates
are at zeroth order, using many simplifications.  A full calculation
of the nuclear burning in the multi-dimensional, asymmetric explosions
is needed to determine which aspects of these results will remain
true.  Still, it is interesting to note that the trends we see for
$^{44}$Ti and $^{26}$Al in these 1D models are contrary to those
reported by other authors in the literature (Nagataki 2000 report
increasing yields for $^{44}$Ti with larger explosion energy and
Limongi \& Chieffi 2005 report constant yields for for $^{26}$Al with
increased explosion energy.)

For our calculations, which focus on the $\gamma$-ray emission, the most
important yield is that of $^{56}$Ni and its decay product: $^{56}$Co.
For most of our calculations, we assume that the $^{56}$Ni remains
unchanged from the predictions of our progenitor model.  But to test
the extreme possible effect of our asymmetries on the $\gamma$-ray line
profile, we have also included two simulations (f2th50enh, f3th40enh)
that enhance the $^{56}$Ni within the energetic lobe by a factor of
two (Table 1).  This corresponds to a total increase in $^{56}$Ni by
roughly 0.02,0.2M$_\odot$ for the $\Theta=20,40$ models respectively.
We have added this $^{56}$Ni from the inside out (to mimic the trends
we see in our 1-dimensional calculations).  The effect of this
enhancement will be discussed in more detail in \S \ref{sec:hesc}.

For weaker explosions, even the stable isotope yields can vary by over
an order of magnitude.  Figure \ref{fig:abund} (bottom) shows the same
abundance fractions with 1-dimensional explosion energies of $0.1
\times 10^{51}$\,ergs (black) where a lot of fallback occurs, $6.5
\times 10^{51}$\,ergs (blue), and the 90\%,10\% linear combination
(red).  Here the small, energetic outburst drastically alters the
predicted yields of the star.  Indeed, in this scenario, even though
we would expect considerable fallback, the yield would be within a
factor of a few for most elements with respect to normal supernova.  
We postpone study of these weak explosions to a future paper.

\section{High Energy Spectral Calculations}
\label{sec:hesc}

We now turn to studying the effects these hydrodynamic asymmetries
have on the $\gamma$-ray spectra and fluxes.  In particular, we can
test to see if one-sided explosion asymmetries can produce redshifts
in the $\gamma$-line profiles.  We concentrate our $\gamma$-transport
simulation efforts primarily on the ejecta structures at a single time
(t = 1 yr)\footnote{We present a few simulations at earlier and later
  times to qualitatively investigate time evolution, though we defer a
  more rigorous study to a future paper.}, and investigate trends in
the line centroid shift with viewing angle.  As we mentioned before,
the progenitor star used in these simulations differs appreciably from
the best-fit progenitor to SN~1987A.  This precludes us from directly
comparing our theoretical line shifts to the $\gamma$-line
observations of Tueller et al. (1990) and the observed [FeII]
forbidden lines from Spyromilio et al. (1990) and Haas et al. (1990).

\subsection{Numerical Schemes}

As with the previous simulations, our input models of the supernova
ejecta (element abundances, density and velocities) are taken from the
f2th20 and f3th40 SPH explosion simulations described above and mapped
onto a 140 $\times$ 140 $\times$ 140 Cartesian grid.  Escaping photons
were tallied into 250 coarse energy bins, with finer binning at the
decay line energies to provide line profile information. The emergent
photons were also tallied into 11 angular bins ( $\Delta \theta$ =
10$^\circ$) .  While the input data is not of an axisymmetric nature,
the set of angles chosen are fairly representative of the overall
structure.

Roughly 5$\times$10$^9$ Monte Carlo photon bundles were generated for
each input model, in proportion to the mass of radioactive material
distributed throughout the ejecta.  For these models, photoelectric
and pair production opacities were calculated for the elements H, He,
C, O, Ne, Mg, Si, Fe, Co, and Ni.  These elements correspond to those
used in the nuclear network by Timmes, Hoffman \& Woosley (2000),
which has been incorporated into the SPH code but was turned off in
these calculations for computational efficiency.  This reduction in
the number of elements treated (relative to HFW03) manifests itself 
as variations in the spectral turnover at low energies, but does
not affect the Compton scattered continuum as $Y_e,$ and hence the
electron density, is the same.  As mentioned above, HFW03 used a 
grid of central SPH particles to represent the central neutron star.
The treatment of particle composition when mapping to the Monte
Carlo grid led to an overproduction of $^{56}$Ni by roughly 35\%.
The simulations presented here replace the central SPH particles
with an external gravitational force and do not suffer from the
nickel overproduction.  In all other aspects, the simulations 
remain the same as in HFW03.

\subsection{Hard X-ray and Gamma-ray Spectrum}

Without the benefit of a progenitor and explosion model tuned
specifically to SN~1987A, it is difficult to draw meaningful
conclusions regarding the continuum portion of the high energy
spectrum.  The hard X-ray continuum was detected earlier than theory
predicted based on spherically symmetric (i.e. unmixed explosion
models.)  Multi-dimensional explosions, capable of modeling the mixing
instabilities, have been simulated using a stellar progenitor
appropriate for SN~1987A (Kifonidis et al. 2003; Nagataki 2000; Herant
\& Benz 1992).  However, the nickel distributions from these globally
symmetric explosion models were not broad enough to match the first
order comparison against IR line profiles, so further modeling of the
high energy spectrum was not performed.  This work tells us that
additional mixing (possibly arising from a global asymmetry) is needed
to match the broad lines, but it tells us nothing regarding the effect
a global asymmetry should have on the continuum.  The high energy
spectra from a subset of our explosion simulations (using a generic
red supergiant SN progenitor) are shown in Figure~\ref{fig:hardx},
where photon flux (phot/s/MeV/cm$^2$) is plotted logarithmically
across the energy range investigated with these simulations (0.3~keV -
4~MeV).  It's clear that the asymmetric explosions produce a lower
continuum relative to the symmetric explosion model, and that the time
evolution of models with different input asymmetries is qualitatively
different.  Previous work has judged its mixing successes and
failures based on the spatial distribution of nickel alone, however,
these results suggest that the hard X-ray emission can vary widely
with asymmetry type and high energy calculations are necessary for
determining the adequacy of any SN~1987A mixing calculation.

\subsection{Gamma-ray Line Profiles}

Figures \ref{fig:f3th40_anggam}~-~\ref{fig:symmgam} show the
847~keV $\gamma$-ray line profiles for our suite of single lobe explosion
models.  The plots are the same layout and represent the same viewing
angles as in Figures \ref{fig:f3th40_nidist} and
\ref{fig:f2th20_nidist}.  These figures demonstrate that redshifted
line profiles {\it are} attainable given the types of asymmetries
assumed here.  However, it is clear by comparing to Figures
\ref{fig:f3th40_nidist} and \ref{fig:f2th20_nidist}, for models f3th40 and 
f2th20, that the line
profiles do not reflect the entirety of the underlying nickel
distribution.  Indeed, as argued in HFW03, the high
energy emission can be understood by assuming that the emission at all
viewing angles is dominated by the extremities of the nickel
distribution (which has mostly decayed to cobalt at this epoch.)

Each energy bin in the Doppler broadened
profile can be mapped to a spatial location in the ejecta.  This is
due to the homologous nature of the expansion, in which line of sight
velocity is proportional to line of sight distance from the mid-plane
of the explosion.  Keeping this in mind, one can use the 
structure of the cobalt contour plots
in Figures \ref{fig:f3th40_anggam}~-~\ref{fig:symmgam} to
understand the line profiles plotted in the surrounding panels.
Each energy bin in the line profile corresponds to emission from 
all the cobalt located along a line perpendicular to the viewing
angle vector (black line drawn from explosion center to line profile
panel.)  

In particular, for the f3th40 model, the base of the antenna-like
structures is the primary emission site for $\gamma$-rays that escape
at all angles.  For the $\theta$=0 direction, this leads to a blue
shifted line with a long blue wing contributed by photons from the
antenna structures themselves.  The $\theta$=90 direction shows two
separate peaks with emission shoulders to the red and blue.  The peaks
arise from the structure just at the split of the two antennae, with
the near side giving rise to the slightly blueshifted peak and the
far side making up the slightly redshifted peak.  The shoulders on
these peaks arise from the emission of the antenna material.  The same
correspondence between the outlying cobalt distribution and line
profile structure can be made for the f2th20 model.  In each panel
there is a small contribution from the mildly blueshifted material at
the outer edge of the spherical structure in the center.  However, for
the viewing angles with a clear view to the cylindrical structure
along the z-axis, the profiles are dominated by that emission.  In
particular, the $\theta$=0 direction shows a strongly blueshifted
peak from the spray at the end of the cylinder and a shallow redward
wing from the emission of the cylinder itself.  In the $\theta$=90
direction, the high z-velocity material in the cylinder has a low
line of sight velocity and gives an emission peak at roughly the rest
frequency.  However, this profile does show extended red and blue
wings arising from the spray at the end of the cylindrical structure.
The simulations with enhanced nickel synthesis, show little difference
from the standard simulations, primarily because the enhancement affects
the inner nickel distribution most strongly, while the $\gamma$-ray lines
are most sensitive to the outwardly mixed nickel.

While it is encouraging that such a line shape can be reproduced at
all, it is important to note that in both models the redshifted line
shape comes at the expense of line flux.  We see significant redshifts
because we see only a fraction of the radioactive emission which has
been mixed out to low densities in the ejecta (and thus low
opacities.)  Indeed, the integrated flux for any viewing angle in the
847~keV line represents only a few percent of the total flux from the
0.24~M\sun of nickel ejected in these models.  The $\gamma$-line
observations of SN~1987A implied 20\% escape fraction at roughly 400
days (Tueller et al. 1990), however the explosion energy for SN~1987A
was likely larger than that of our simulations (Woosley 1988).  As a 
result, we would expect the timescales for uncovering the radioactive
nickel to be different in SN~1987A so as to account for the larger
escape fraction.  However, for the models considered here, a larger escape
fraction would mean a more significant contribution from the
spherically distributed nickel deeper in the ejecta, possibly washing
out the redshift in the observed profile.

In order to probe the effect a larger escape fraction might have on
the overall $\gamma$-ray lineshifts, we look to later times in the
evolution of the f2th20 and f3th40 explosion models.
Figures~\ref{fig:f3th40_anggam}~and~\ref{fig:f2th20_anggam} show the
847~keV $^{56}$Co line profiles for t~=~250 and 600~days overplotted
with the profiles at 1 year.  There is a clear evolution with time of
the line centroid.  As the ejecta expand, the ``observed'' line
profile shifts toward the decay rest energy regardless of viewing
angle.  So we find that these particular models seem to have
difficulty reproducing the time independence of the redshifted line
profiles in SN~1987A.  However, these simulations also suggest that
the general nature of the extended nickel distribution varies
sensitively with assumed velocity asymmetry.  One can imagine that it
is possible to reconstruct the specifics of the SN~1987A observations
with an appropriately distributed ``spray'' of radioactive nickel
products.  There remain many ``ifs'' regarding the ability of this
mechanism to match the line fluxes and time evolution of the
$\gamma$-ray lines of SN~1987A.  However, the outlook is good that
this may be the solution we have been looking for and a serious effort
to match the SN~1987A data specifically using asymmetries of this
nature is definitely warranted.  For the specific explosion parameters
assumed here, we can already conclude that the asymmetry in the
explosion manifests itself most clearly in the ``spray'' of heavy
elements along the enhanced explosion lobe.  It is the detailed
structure of this nickel ``spray'' that is directly probed by the
$\gamma$-ray line profiles at this epoch, thus probing the material
most likely to indicate a break in global symmetry.

\section{Summary}

The hydrodynamic simulations investigated here probe only a small
portion of the large parameter space for explosion asymmetries.  From
this limited sample we can say that the outward mixing of heavy
elements in these single lobe explosion models is more extended than
the bipolar explosion simulations of HFW03.  The differences in
adopted angular profile of the imposed velocity asymmetry is likely
responsible for this enhancement, rather than the difference between
bipolar and single lobe explosion geometry.  The overall morphologies
in the high {\it f} explosions (f5th20, f3th20 and f3th40) are
particularly reminiscent of the Cassiopeia A supernova remnant, in
that the heavy element distribution shows a clear ``jet'' blowing out
of the star.  The persistence of such a density asymmetry through the
extended envelope of our RSG progenitor is promising, primarily for
matching the polarization observations in Type~II SNe.  It is also
heartening that remnant morphologies like Cas~A can be achieved
through explosion asymmetries, though this is less crucial since it is
entirely possible that stellar wind profiles and binary interactions
have a larger affect on the remnant structure than the explosion
mechanism. 

The high energy transport simulations, which are calculated as a
post-process on the hydrodynamic models, verify that redshifted line
profiles are attainable for $\gamma$-ray decay emission in single lobe
explosion asymmetries.  However, the redshifted emission is primarily
attributed to the ``spray'' regions of enhanced outward mixing of cobalt.
This means that the specifics of the $\gamma$-ray line profiles are
strongly tied to the structure of the hydrodynamic mixing and do not probe
the entirety of the nickel distribution.  This is extremely fortunate, as
it is this outlying material that contains the most information regarding
the initial velocity asymmetry.  In this way, the combined sensitivity of
the $\gamma$-rays to the ``spray'' material, and the ``spray'' material to
the underlying velocity structure, make the $\gamma$-rays an ideal probe of
the explosion mechanism itself.  This begs the question, what is the 
likelihood of detecting the $\gamma$-ray line emission from core-collapse 
supernovae with current and future $\gamma$-ray instruments?  Observations at
these high energies require space observatories, or at the very least,
high-altitude balloon missions.  The current state-of-the-art for
$\gamma$-ray observations is the International Gamma-Ray Astrophysics
Laboratory (INTEGRAL, operated by European Space Agency).  A caveat to
keep in mind is that the nickel mass synthesized in this 15\,M\sun \
model is roughly 2 times larger than the mean observed value
($\sim$0.13\,M\sun; Hamuy 2003) for core-collapse SN explosions.
However, as mentioned earlier, the explosion energy was roughly half
that of a normal supernova.  These two effects serve to balance one
another, though it is difficult to determine the exact effect as
nickel distributions and optical depth profiles are not simply
described.

At energies around 1~MeV, INTEGRAL will have a spectral resolution of
2~keV and a narrow line sensitivity ($3~\sigma$ in 10$^{6}$~seconds)
of $\sim$5$\times$10$^{-6}$~phot~s$^{-1}$~cm$^{-2}$ (Hermsen \&
Winkler 2002).  Our model lines are about 5 times broader than this
resolution element, so the sensitivity for detecting them is worse by
roughly $\sqrt{5}$.  Using these specifications, INTEGRAL would be
able to detect the $^{56}$Co lines from the single lobe explosion
models (looking along the explosion lobe) at a distance of roughly
850~kpc and 300~kpc for the f3th40 and f2th20 models respectively.
Table~\ref{tab:fluxes} shows the luminosities for various continuum
energy bands and line energies from the four asymmetric models studied here
(f2th20, f3th20, f5th20 and f3th40).  At these distances (less than a
Mpc), the occurrence rate for core-collapse supernovae is essentially
the rate for a Galactic event (roughly 1-2 per century; Cappellaro et
al. 1993).  For a Galactic supernova event, INTEGRAL would be able to
measure not only line flux but line profile information, allowing the
full diagnostic potential of the $\gamma$-rays to be tapped.

While the signatures of asymmetry in any one of our models do not
exactly match the specific observations of SN~1987A or Cas~A SNR, the
set of simulations in HWF03 and this work seem to span the range of
observed features.  For our most extreme velocity asymmetries, we see
``jet''-like morphologies in the ejecta density plots and iron group
elements mixed to the outer edge of the SN ejecta.  Redshifted line
profiles can be seen along certain viewing angles in the single lobe
explosions, as can broadened line profiles.  In addition, the
discontinuous velocity asymmetry in these single lobe explosions
produces high velocity nickel clumps, an important ingredient in the
explanation of SN~1987A's Bochum event.  Finally, the single lobe
explosions seem to result in dimmer continuum emission relative to the
symmetrically mixed explosion models.  We note that the smoother,
bipolar explosion model in HFW03 shows an enhancement in continuum
emission and results in even broader line profiles than are seen in
these single lobe explosion models.

An important ramification of these results is that features associated
with ejecta asymmetry are reproduced naturally within the context of
the standard neutrino driven supernova mechanism.  Our input
parameters were motivated by the core collapse simulations of Scheck
et al. (2004)\footnote{Likewise, the input asymmetry in HFW03 was
motivated by core collapse calculations of Fryer \& Heger (2000).}, in
which neutrino driven convective motions are responsible for the
explosion asymmetry\footnote{A caveat to keep in mind is that the
asymmetry inferred from the collapse simulations is not imposed onto
the ejecta until the shock has nearly reached the hydrogen layer.
However, the shock primarily accelerates until it reaches this layer
and any velocity asymmetry is likely to persist until this point.}.
The simulations presented in this paper, and in HWF03, suggest that a
combination of the bimodal and unipolar asymmetries may well explain
all the observable features of SN~1987A or Cas~A SNR.  {\it No
jet-induced explosion need be invoked for these, and indeed, most
supernova explosions.}  Without knowing it, Fryer \& Warren (2004)
already produced such a combination in the explosion of a rotating
star (model SN15B).  Obviously, a convincing demonstration of this
awaits the simulation of a progenitor model specific to SN~1987A or
Cas~A SNR and a more comprehensive study of the asymmetries.  But the
pieces are now on the table and we plan to continue with more object
specific studies to convincingly solve this puzzle.

\acknowledgements This work was funded under the auspices of the U.S.\ 
Dept.\ of Energy, and supported by its contract W-7405-ENG-36 to Los
Alamos National Laboratory, by a DOE SciDAC grant number
DE-FC02-01ER41176 and by NASA Grant SWIF03-0047-0037.  The simulations
were conducted on the Space Simulator at Los Alamos National
Laboratory.

\clearpage

\begin{deluxetable}{lcccc}
\tablecaption{Summary of Simulations}

\tablehead{ \colhead{Model} & \colhead{$\Theta$} 
& \colhead{f} &\colhead{$v_{\rm kick} ({\rm km s^{-1}})$}} 

\startdata
symm   & 0  & 1 & 0 \\
f2th20 & 20 & 2 & 320 \\
f2th20enh\tablenotemark{a} & 20 & 2 & 320 \\
f3th20 & 20 & 3 & 600 \\
f5th20 & 20 & 5 & 1010 \\
f3th40 & 40 & 3 & 1690 \\
f3th40enh\tablenotemark{a} & 40 & 3 & 1690 \\
\enddata

\tablenotetext{a}{Abundances enhanced by increasing the total $^{56}$Ni
  abundance by $\sim$100\% inside the cone.}

\end{deluxetable}

\clearpage

\begin{deluxetable}{lccccc}
\label{tab:fluxes}

\tablecaption{Line and Continuum Fluxes}

\tablehead{ \colhead{Model} & \colhead{3-30~keV} & \colhead{30-100~keV} & \colhead{100-500~keV} & \colhead{847~keV Line} & \colhead{1238~keV Line} \\ 
\colhead{}    & \colhead{(erg~s$^{-1}$)} & 
\colhead{(erg~s$^{-1}$)} & \colhead{(erg~s$^{-1}$)} &
\colhead{($\gamma$~s$^{-1}$)} & \colhead{($\gamma$~s$^{-1}$)}}

\startdata
f2th20              & 6.133(0.080)$\times$10$^{37}$
                    & 2.731(0.065)$\times$10$^{38}$
                    & 9.929(0.394)$\times$10$^{38}$
                    & 3.087(0.144)$\times$10$^{44}$
                    & 2.614(0.176)$\times$10$^{44}$\\

f3th20              & 7.626(0.095)$\times$10$^{37}$
                    & 3.182(0.073)$\times$10$^{38}$
                    & 9.848(0.413)$\times$10$^{38}$
                    & 1.078(0.109)$\times$10$^{44}$
                    & 2.348(0.199)$\times$10$^{44}$\\

f5th20              & 5.331(0.101)$\times$10$^{37}$
                    & 2.719(0.088)$\times$10$^{38}$
                    & 8.698(0.496)$\times$10$^{38}$
                    & 0.790(0.112)$\times$10$^{44}$
                    & 1.964(0.224)$\times$10$^{44}$\\

f3th40              & 2.882(0.055)$\times$10$^{37}$
                    & 2.611(0.067)$\times$10$^{38}$
                    & 16.45(0.521)$\times$10$^{38}$
                    & 1.023(0.030)$\times$10$^{45}$
                    & 9.531(0.348)$\times$10$^{44}$\\
\enddata

\end{deluxetable}

\clearpage

\clearpage

\clearpage

\clearpage

\clearpage

\begin{figure}[htp]
\plotone{f5.eps}
\caption{Similar to the previous figure, but for model f2th20 at 
t\,=\,150\,d.  The ejecta have already reached the phase of
homologous expansion, so can be directly compared with the f3th40
results.}
\label{fig:mass_f2th20}
\end{figure}

\clearpage

\begin{figure}[htp]
\plotone{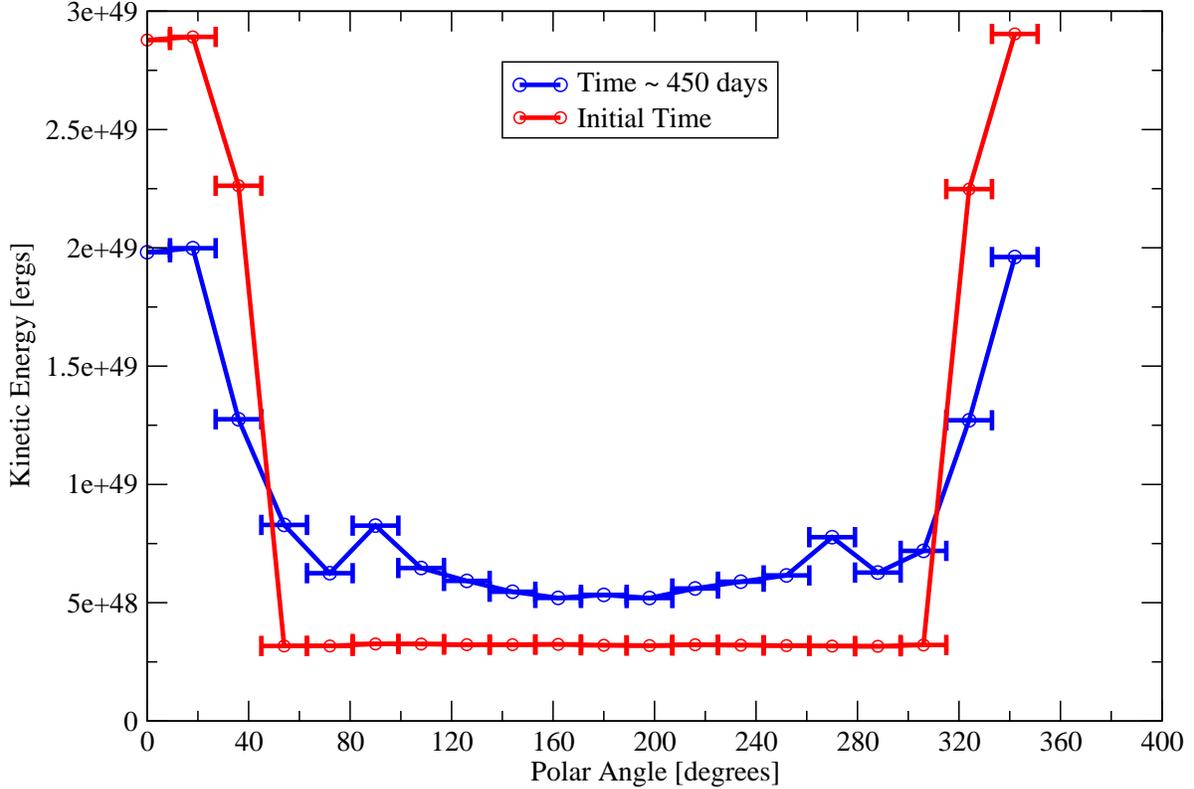}
\caption{Similar to the previous figures, but showing
kinetic energy as a function of polar angle for the f3th40 model.  
Blue line is for the initial time at t~=~100~s and red line is for 
t~=~450~days (long after flow has become homologous.)  It is clear that 
the explosion is spherizing (i.e. energy in the enhanced explosion lobe
is smearing and being shared with the rest of the ejecta.)  Note from the 
previous figures that the
mass in the cone actually increases at later times, suggesting that the
velocity structure is even more spherical than the energy distribution 
shown here.  Still, after having the shock pass through the entire star,
there does still remain some asymmetry in contrast to the complete 
spherization of the bipolar explosion models. }
\label{fig:energy_f3th40}
\end{figure}

\clearpage

\begin{figure}[htp]
\plotone{f7.eps}
\caption{Similar to the previous figure, but for model f2th20 at t\,=\,150\,d.
The ejecta have already reached the phase of homologous expansion, so can be
directly compared with the f3th40 results.}
\label{fig:energy_f2th20}
\end{figure}

\clearpage

\begin{figure}[htp]
\plotone{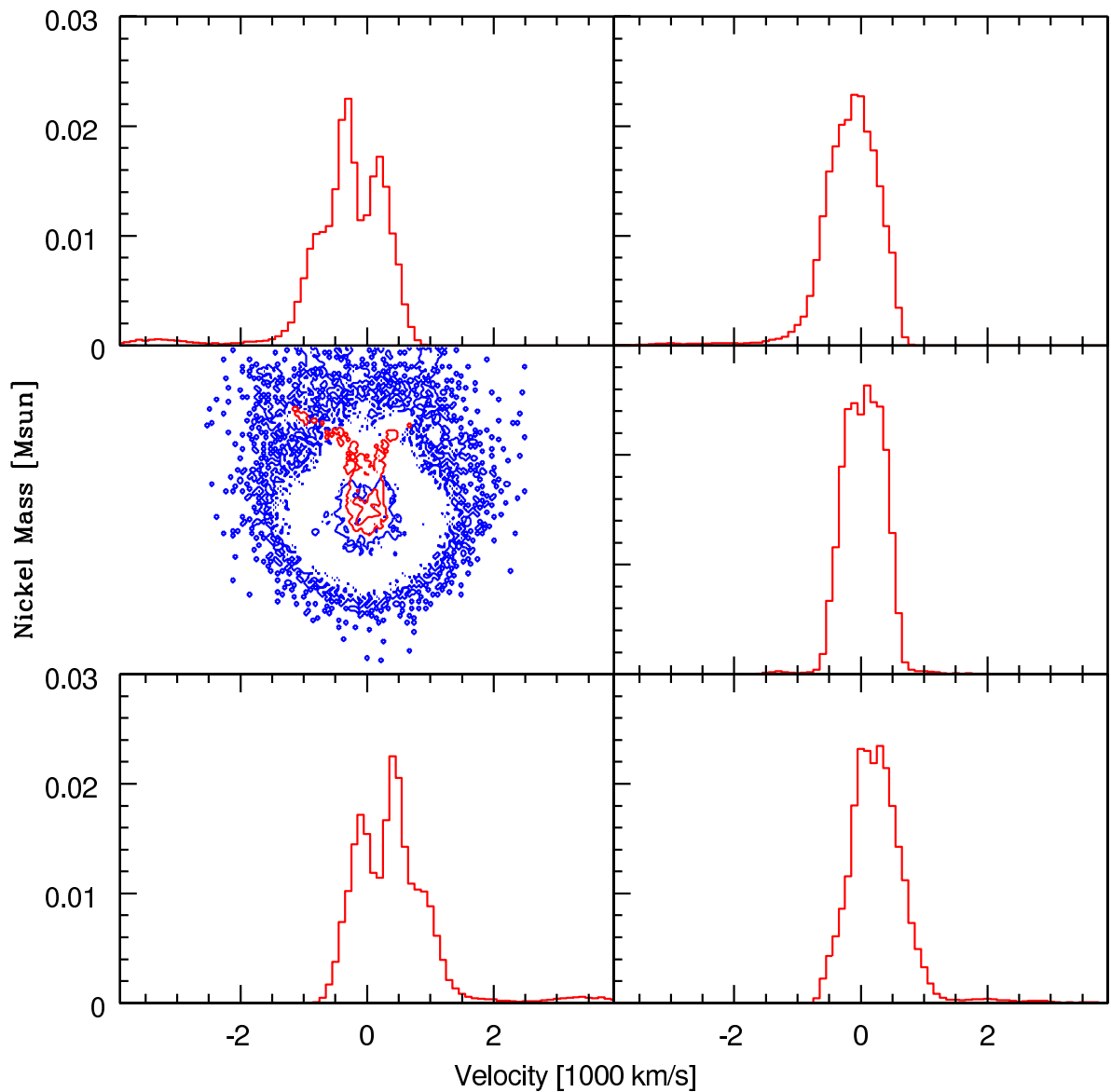}
\caption{Mass of 56-weight elements (e.g. initial $^{56}$Ni mass)
 versus line of sight velocity for a number of viewing angles in model
f3th40 at t~=~365~days.  Central panel on the left shows a contour plot of 
density (blue) and cobalt number density (red).  Due to the homologous
nature of the expansion, these distributions represent the line shapes
one would expect from nickel, cobalt or iron emission in the absence
of significant ionization or opacity effects.}
\label{fig:f3th40_nidist}
\end{figure}

\clearpage

\begin{figure}[htp]
\plotone{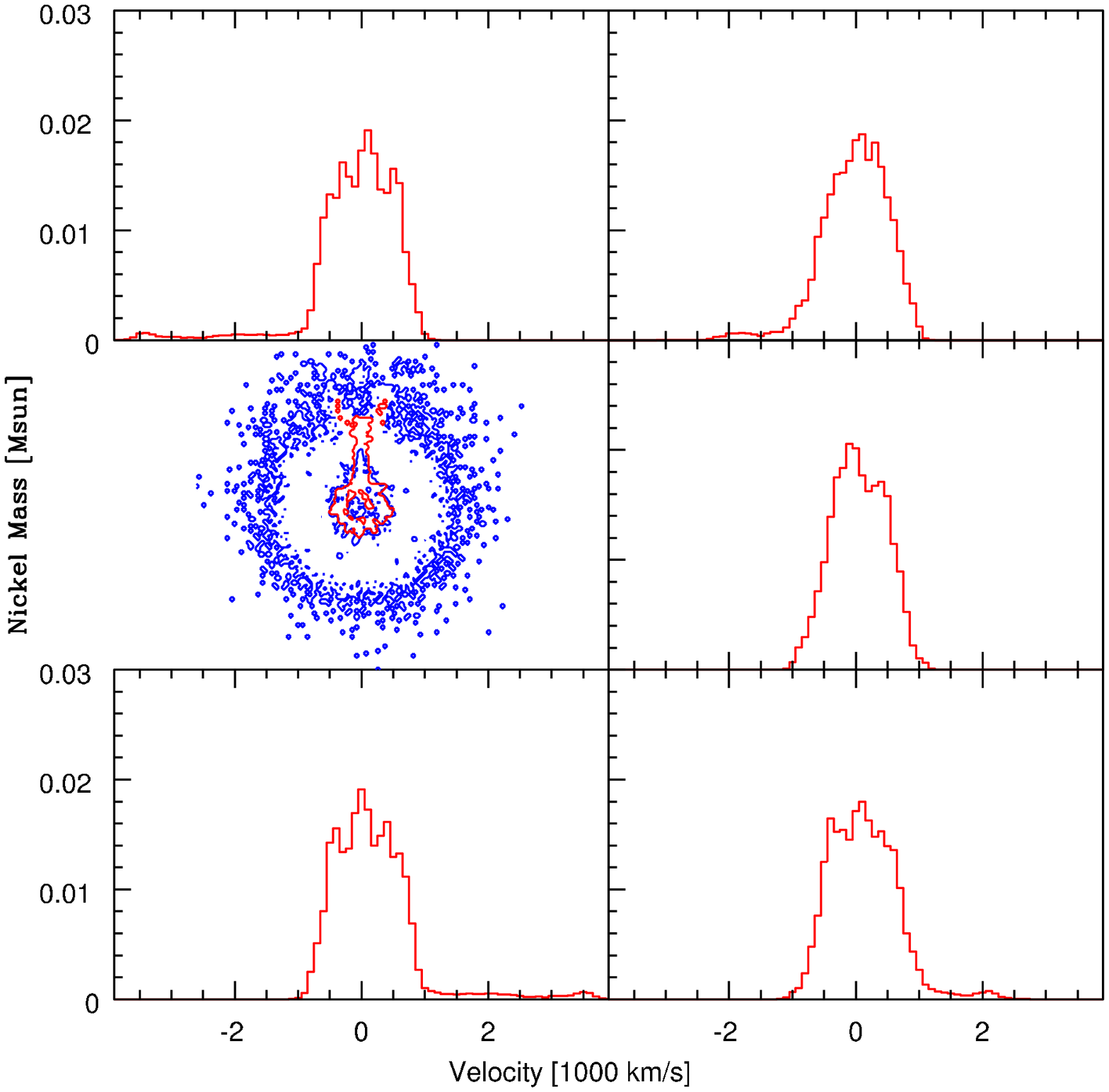}
\caption{Same as previous figure, but for model f2th20.}
\label{fig:f2th20_nidist}
\end{figure}

\clearpage

\begin{figure}[htp]
\plotone{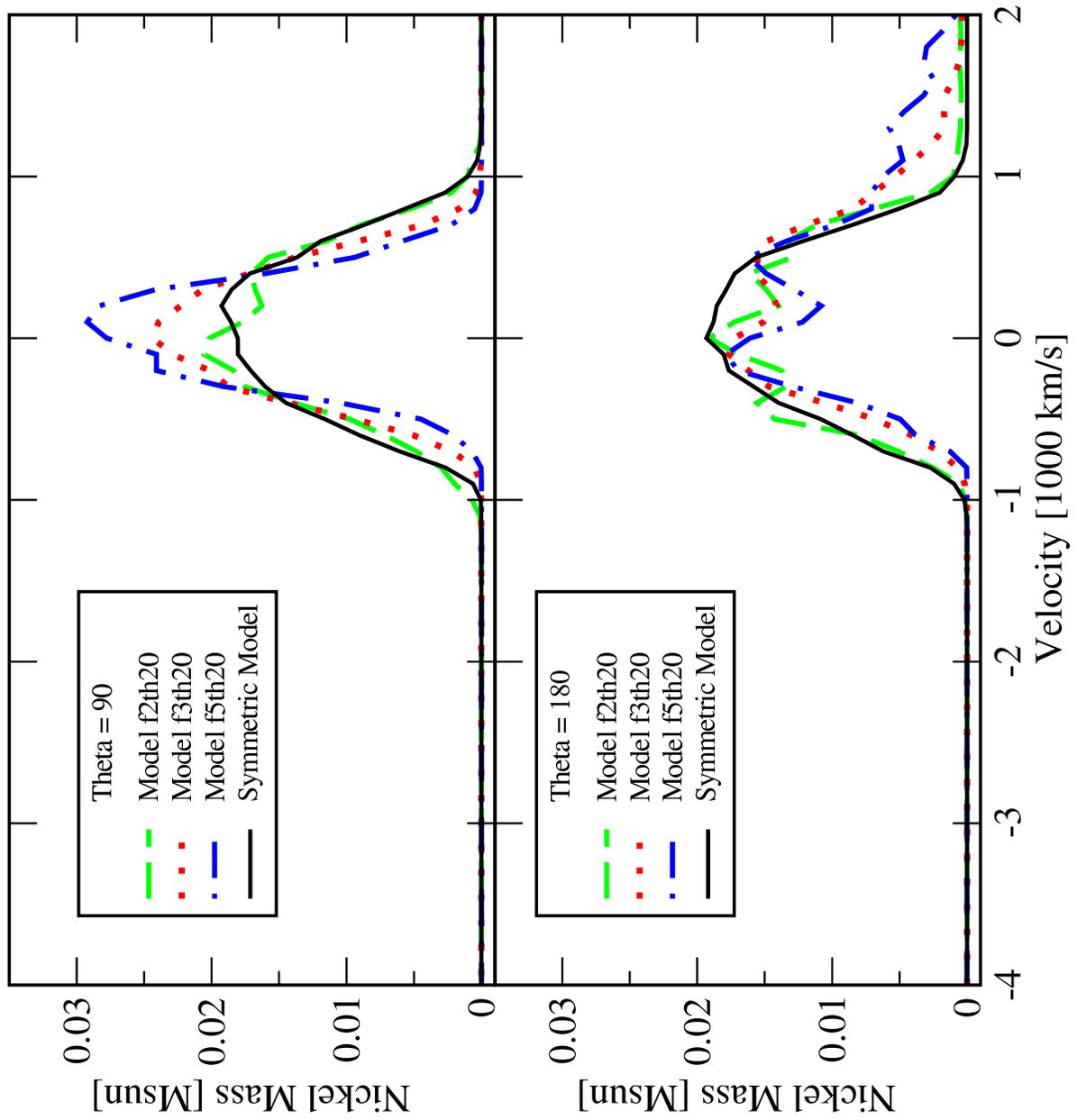}
\caption{Underlying nickel distributions for the fXth20 models, with 
symmetric model distribution plotted for comparison.}
\label{fig:nidist}
\end{figure}

\clearpage

\begin{figure}[htp]
\plottwo{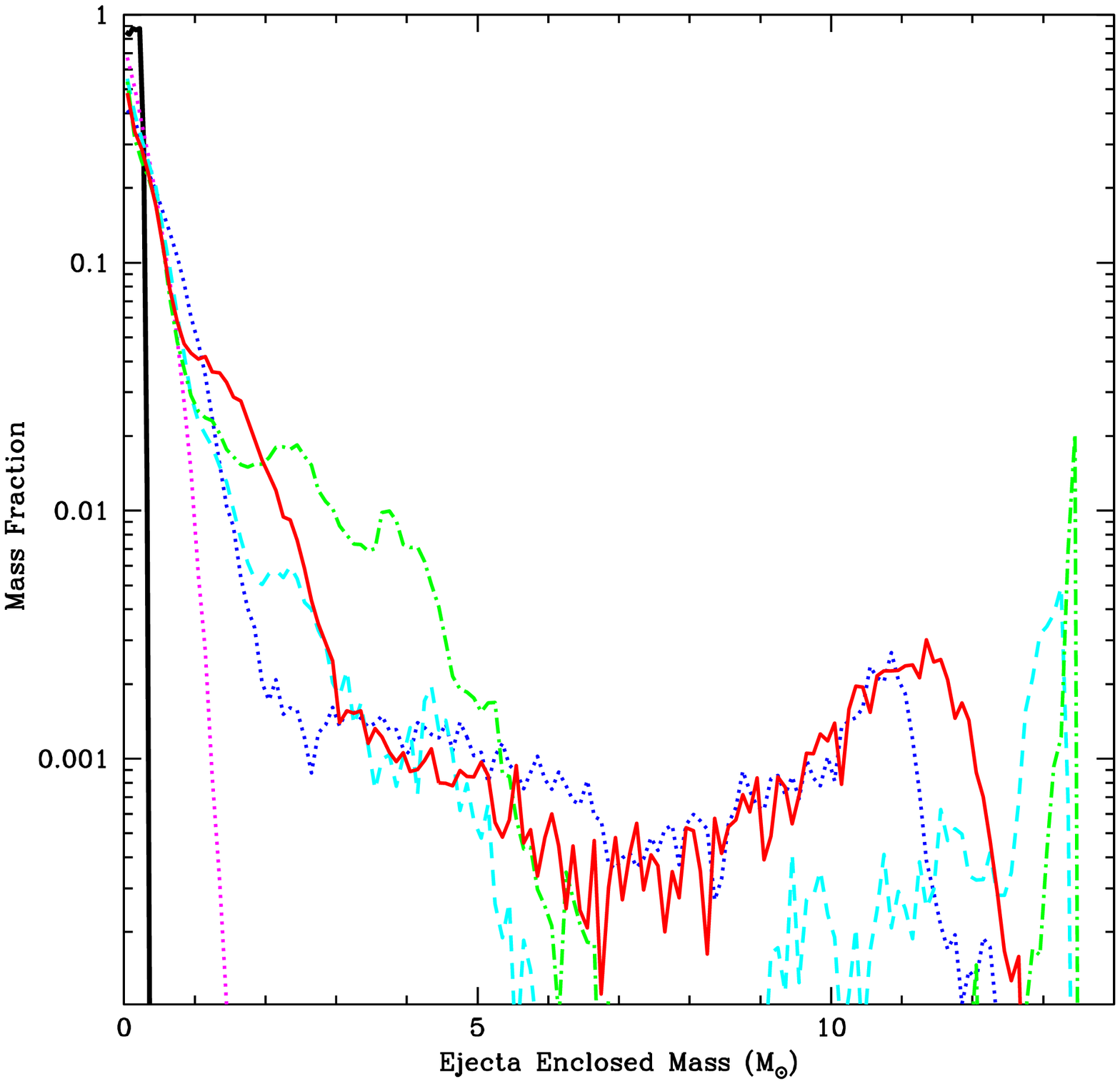}{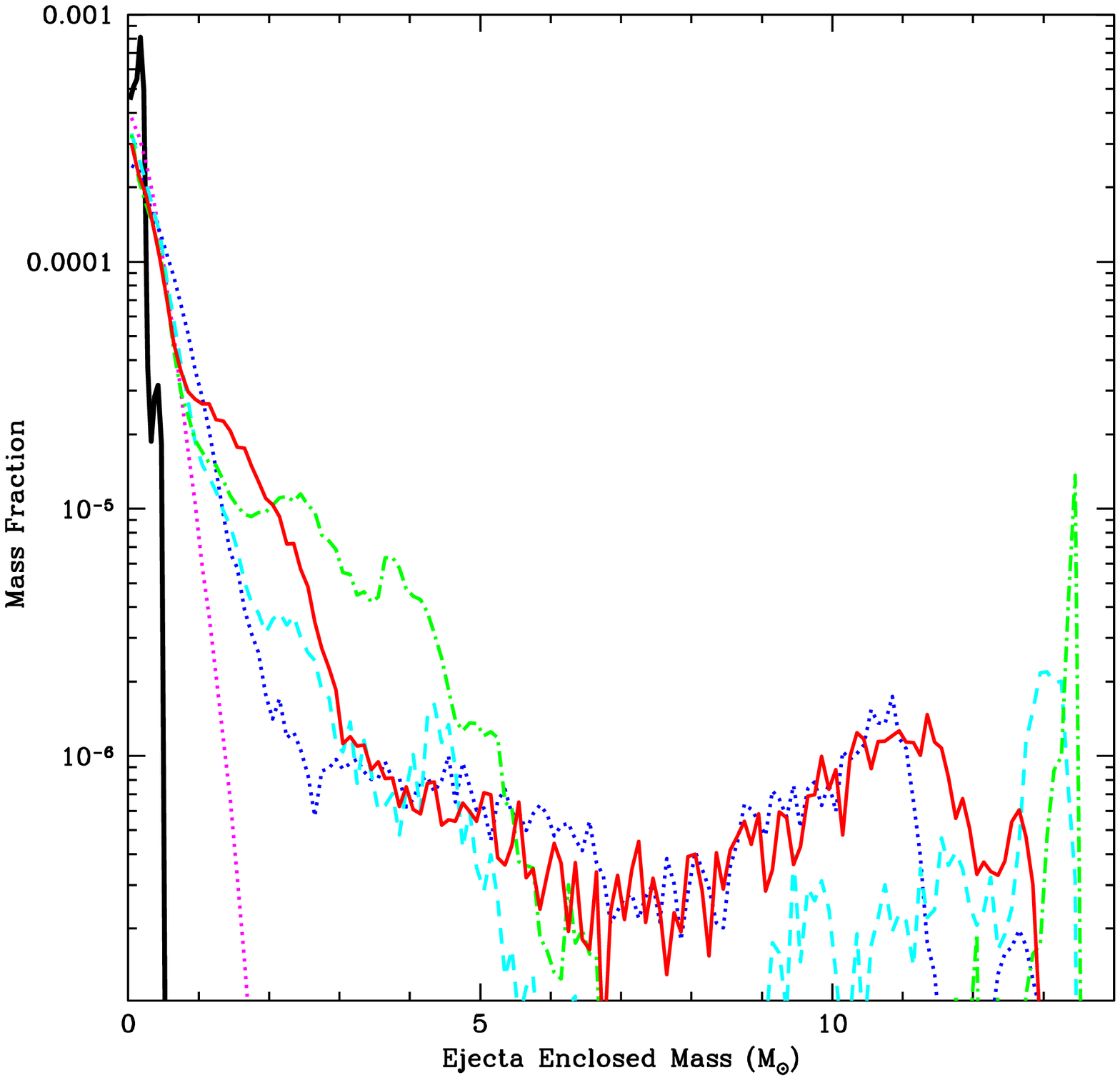}
\caption{$^{56}$Ni mass fraction (left) and $^{44}$Ti mass fraction (bottom)
plotted versus enclosed mass for our suite of explosion models.  Dotted magenta
line is model symm, dotted blue line is model f2th20, dashed cyan line is model
f5th20, dash-dot green line is model f5th20, solid red is model f3th40 and the
nearly vertical line is the initial condition (i.e. the result one would obtain
had no mixing occurred.}
\label{fig:encmass}
\end{figure}

\clearpage

\begin{figure}[htp]
\plotone{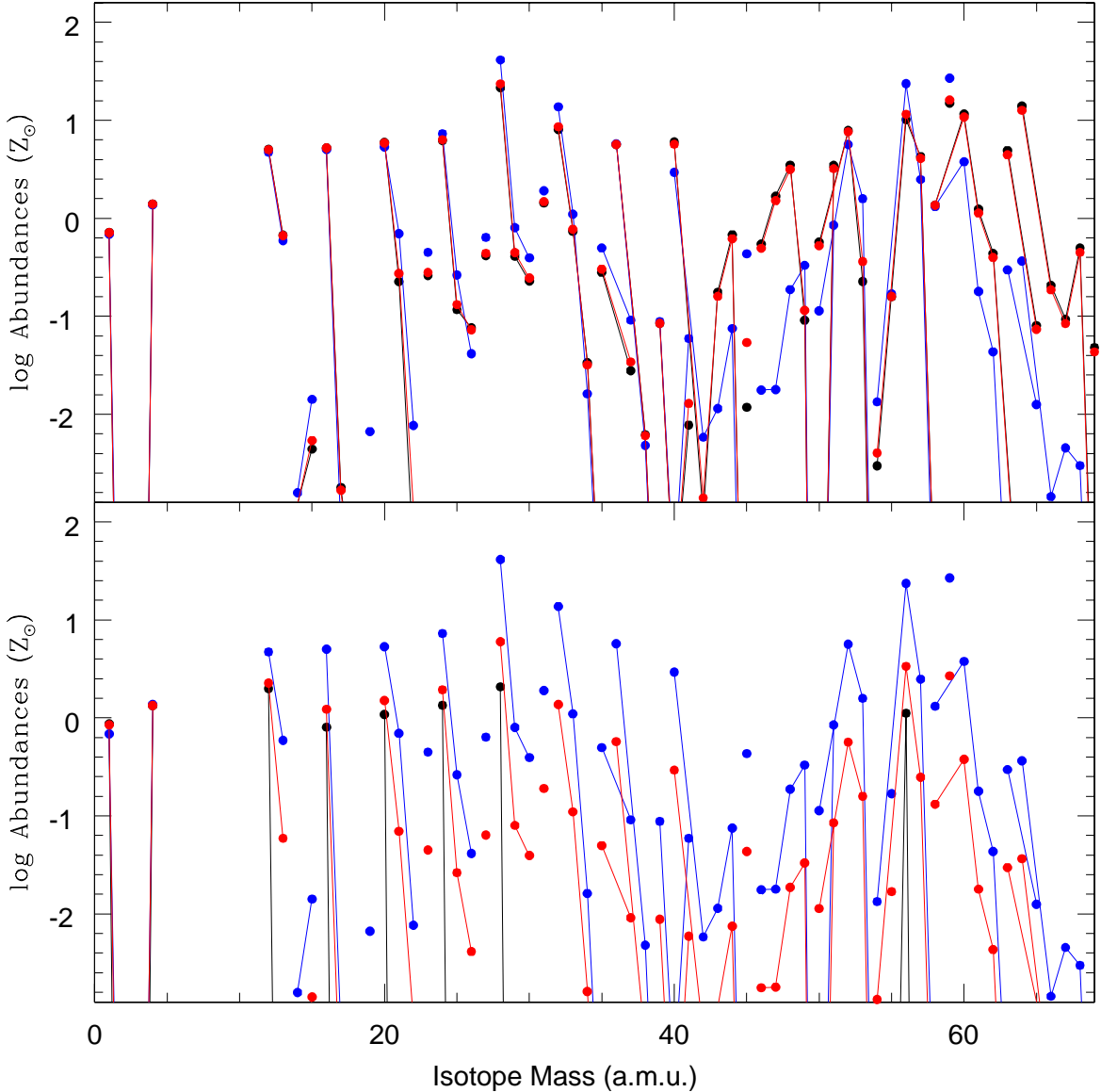}
\caption{Top:  Fractional abundance (in solar units) versus isotope mass for 
  1-dimensional models with $1.35, 6.5 \times 10^{51}$\,ergs (black,
  blue respectively) and for a linear combination (90\%,10\% mix for
  the $1.35, 6.5 \times 10^{51}$\,erg explosions) model.  Note that
  the combined model which presents an extreme result for our current
  models is not too different from a $1.35\times 10^{51}$\,erg
  explosion alone.  Bottom: same set of abundances, but
  substituting a $0.1\times 10^{51}$\,erg explosion for the
  $1.35\times 10^{51}$\,erg explosion.  Although such a result does
  not correspond to our current simulations, it is an example of where
  asymmetric explosions may drastically alter our expectations of the
  nuclear yields.  This example corresponds to an explosion which
  ultimately produces a black hole, though the yields are not too
  different than a normal supernova explosion.  However, be aware that
  these results are very rough estimates, based on very simplistic
  calculations.}
\label{fig:abund}
\end{figure}

\clearpage

\begin{figure}[htp]
\plotone{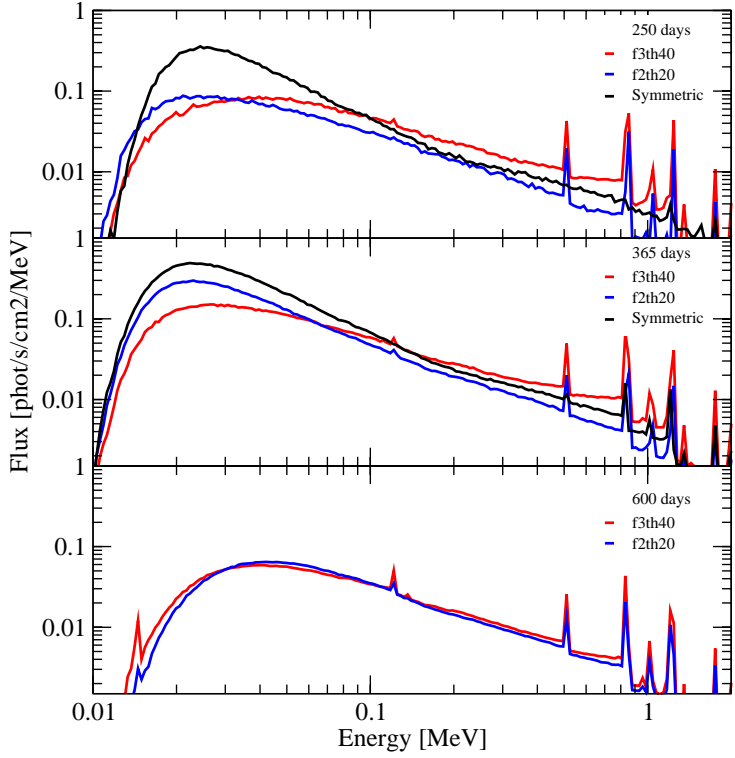}
\caption{Logarithmic plot of total hard X- and $\gamma$-ray spectrum at
  t\,=\,365\,d for models f3th40 (red line), f2th20 (blue line) and
  Symmetric (black line) for 3 different times (t = 250\,d, 365\,d and
  600\,d.)  The flux was calculated assuming a distance of 60~kpc.  The
  f3th40 model evolves more slowly in time, and the Symmetric model
  has a brighter low energy continuum at all epochs.}
\label{fig:hardx}
\end{figure}

\clearpage

\begin{figure}[htp]
\plotone{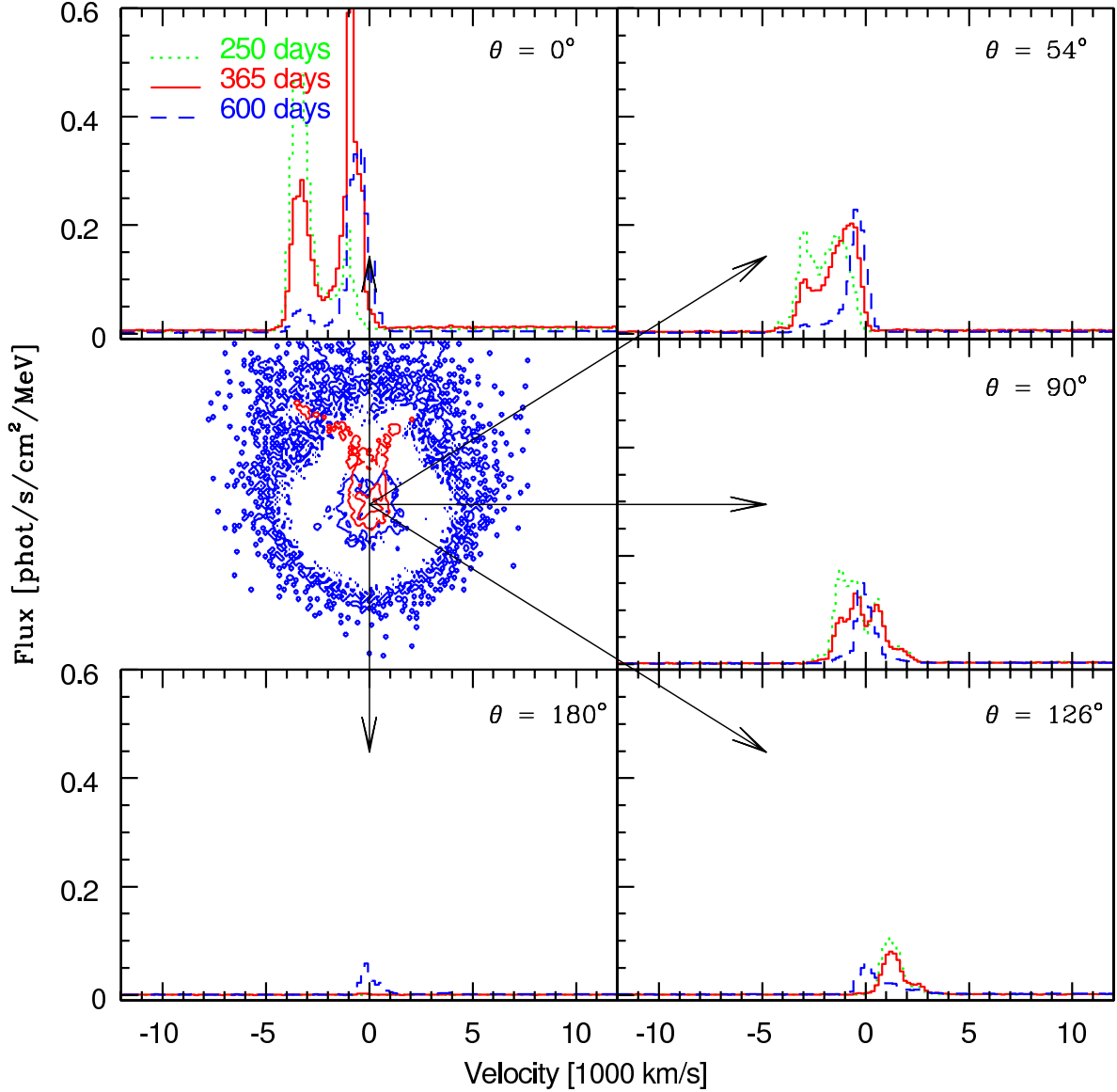}
\caption{Line profiles of the $^{56}$Co 847 keV decay line
for model f3th40 at t~=~365~days.  
Central panel on the left shows a contour plot of 
density (blue) and cobalt number density (red).  Surrounding
panels represent line profiles for the set of viewing
angles depicted by the black vectors overplotted on the density
contours.  The emission in the line profiles arises predominantly
from the cobalt ejected along the enhanced explosion lobe.  Due
to the homologous nature of the ejecta, the 
structure in the lines can be understood by summing this extended 
cobalt material along lines perpendicular to the viewing angle
vectors.  See text for a more in depth discussion.}
\label{fig:f3th40_anggam}
\end{figure}

\clearpage

\begin{figure}[htp]
\plotone{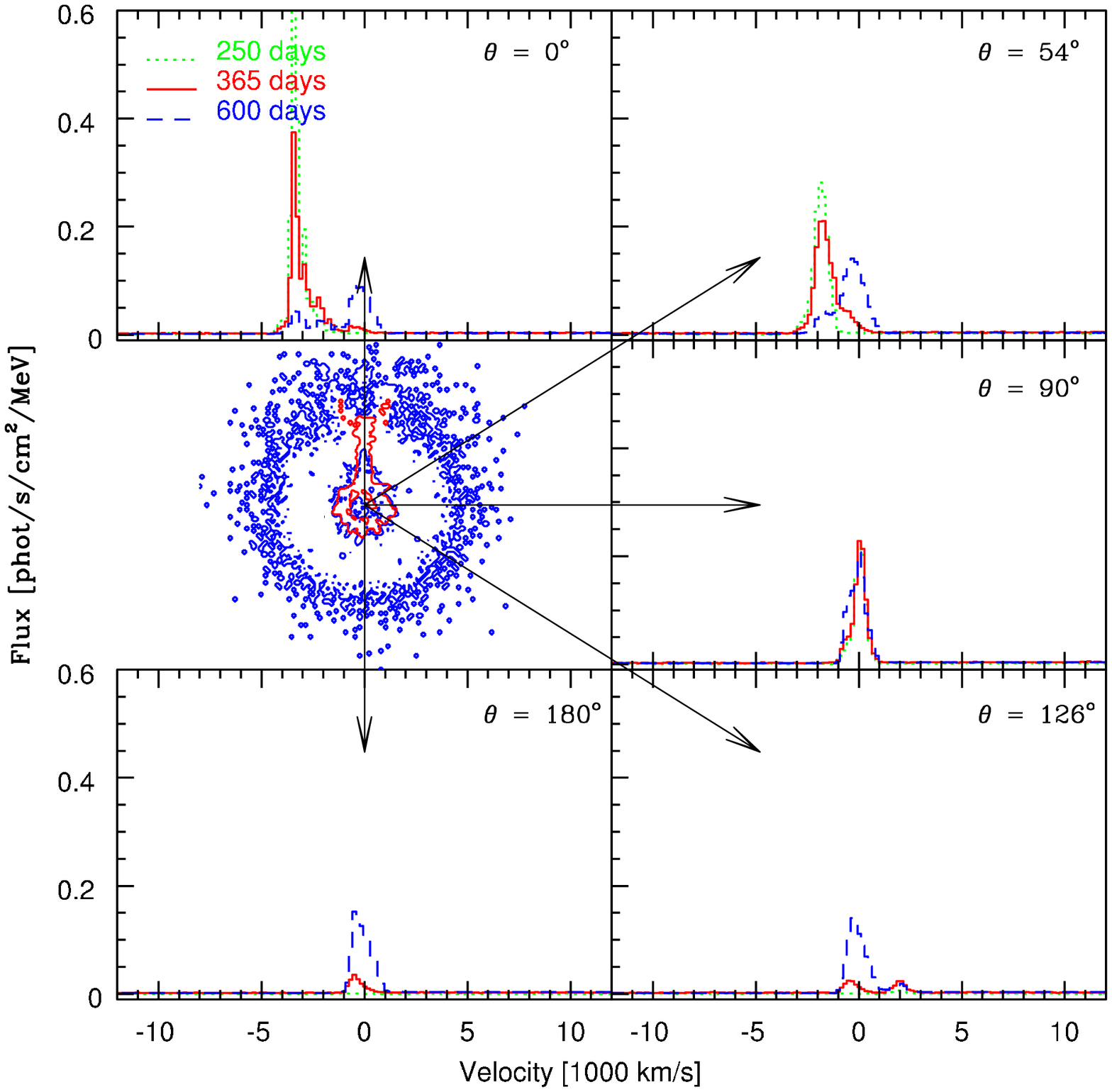}
\caption{Same as previous figure, but for model f2th20.}
\label{fig:f2th20_anggam}
\end{figure}

\clearpage

\begin{figure}[htp]
\plotone{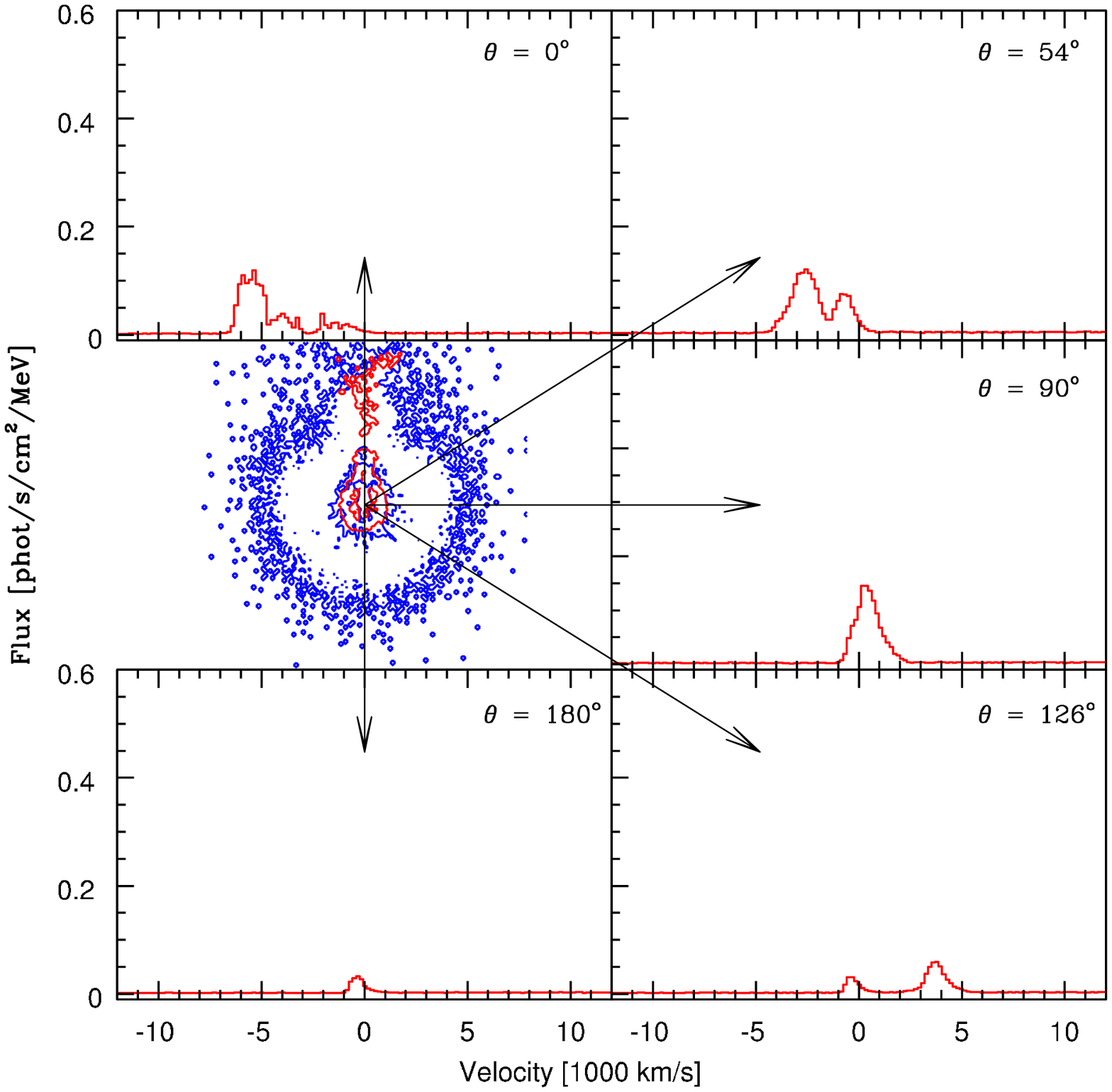}
\caption{Same as previous figure, but for model f3th20.}
\label{fig:f3th20gam}
\end{figure}

\clearpage

\begin{figure}[htp]
\plotone{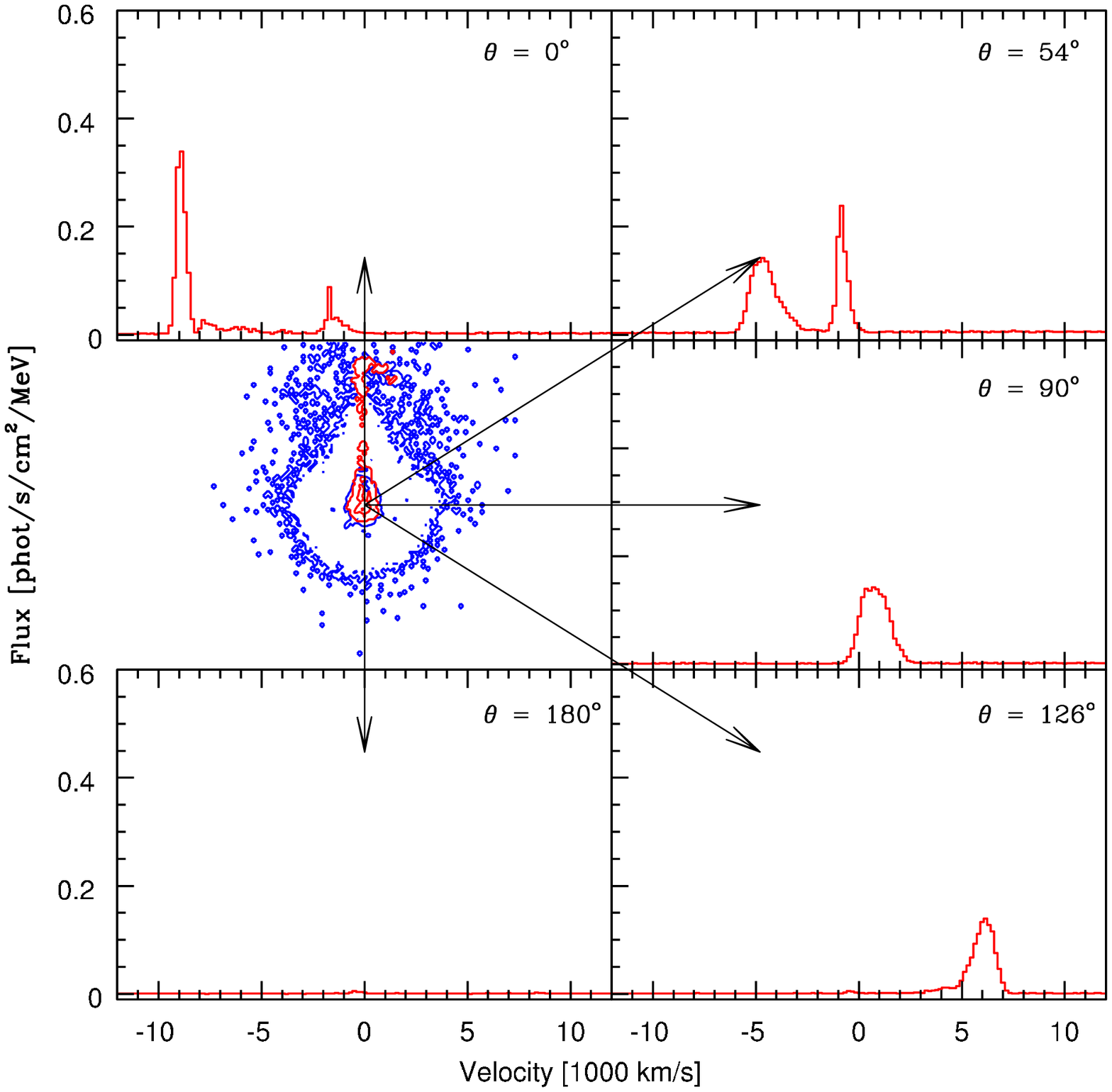}
\caption{Same as previous figure, but for model f5th20.}
\label{fig:f5th20gam}
\end{figure}

\clearpage

\begin{figure}[htp]
\plotone{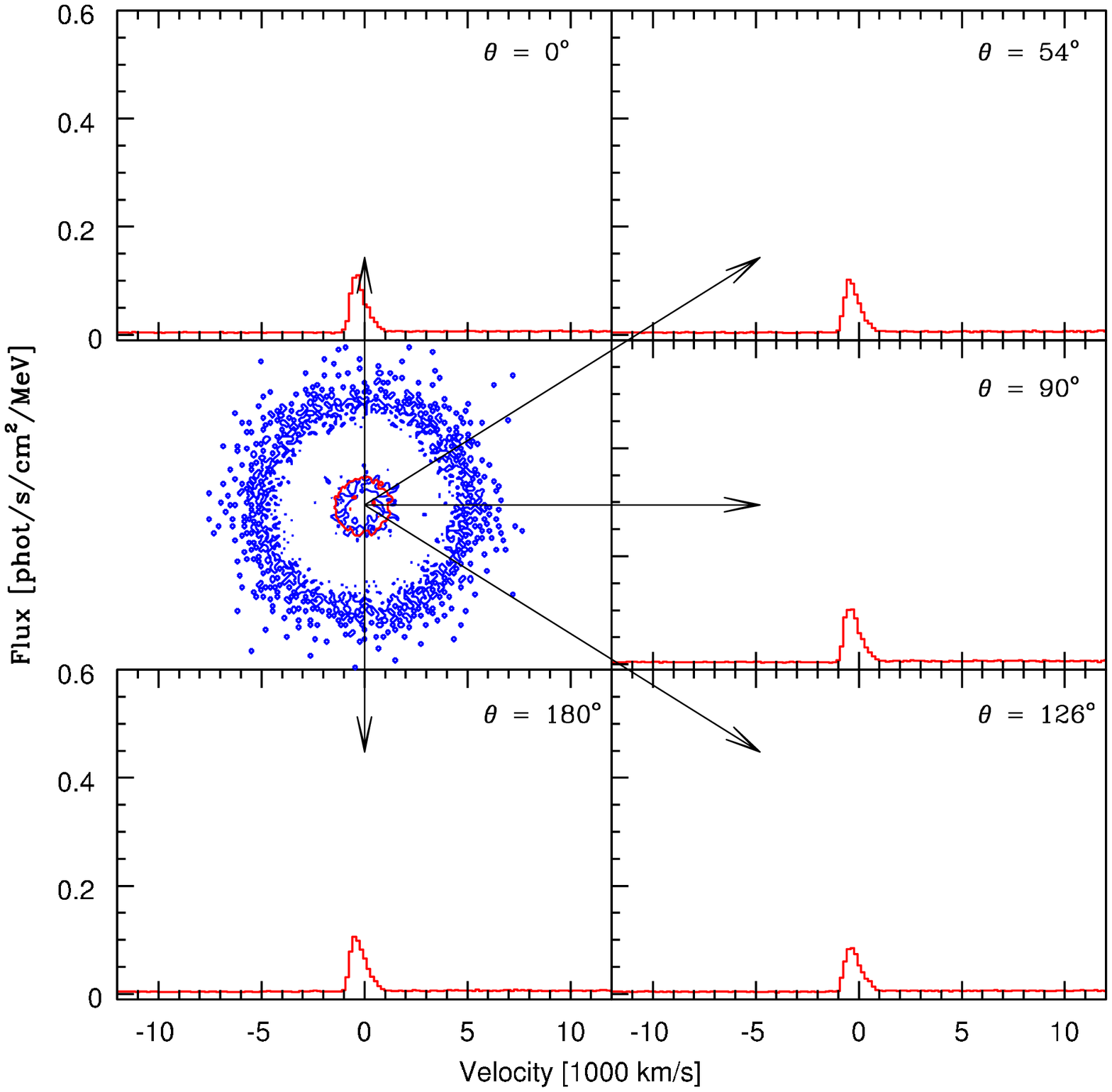}
\caption{Same as previous figure, but for model symm.}
\label{fig:symmgam}
\end{figure}


\clearpage

\begin{thebibliography}{}
  
\bibitem{Arn89b}
Arnett, W. D.,  Bahcall, J. N., Kirshner, R. P., \& Woosley, S. E. 
1989, ARA\&A, 27, 629

\bibitem{Arr99}
Arras, P. \& Lai, D., 1999, ApJ, 519, 745

\bibitem{Arz02}
Arzoumanian, Z., Chernoff, D. \& Cordes, J., 2002, ApJ, 568, 289

\bibitem{Blo03}
Blondin, J., Mezzacappa, A. \& DeMarino, C., 2003, ApJ, 584, 971

\bibitem{cha02}
Chatterjee, S., \& Cordes, J. 2002, ApJ, 575, 407

\bibitem{Dot87}
Dotani, T., Hayashida, K., Inoue, H., Itoh, M., \& Koyama, K. 1987, 
Nature, 330, 230

\bibitem{Fry98}
Fryer, C., Burrows, A. \& Benz, W., 1998, ApJ, 496, 333

\bibitem{Fry00}
Fryer, C. L., \& Heger, A. 2000, ApJ, 541, 1033

\bibitem{Fry02}
Fryer, C., \& Warren, M. 2002, ApJ, 574, 65

\bibitem{Fry04}
Fryer, C., 2004, ApJ, 601L, 175

\bibitem{Gra97}
Grant, K. J. \& Dean, A. J., 1993, A\&AS 97, 211

\bibitem{Haa90}
Haas, M. R., Erickson, E. F., Lord, S. D., Hollenbach, D. J., 
Colgan, S. W. J., \& Burton, M. G. 1990, ApJ, 360, 257

\bibitem{Her92}
Herant, M., \& Benz, W. 1992, ApJ, 387, 294

\bibitem{Heretal94}
Herant, M.,  Benz, W., Hix, W.R., Fryer, C.L. \& Colgate, S.A. 
1994, ApJ, 435, 339

\bibitem{Her95}
Herant, M., 1995, SSRv, 74, 335

\bibitem[Higdon et al.(2004)]{Hig04} Higdon, J., Lingenfelter, R., \&
Rothschild, R. 2004, ApJL, 611, 29

\bibitem{Hof91}
H\"oflich, P., 1991, A\&A, 246, 481

\bibitem{hung03} Hungerford, A.L., Fryer, C.L., \&  Warren, M.S. 2003, ApJ, 594, 390
\bibitem[Hungerford et al. (2004)]{Hun04} Hungerford, A. L., Fryer, C.
  L., Timmes, F. X. \& McGhee, K. , submitted to Nuc. Phys. A

\bibitem[Hungerford et al. (2005)]{Hun05} Hungerford, A. L., Fryer, C.
  L., Timmes, F. X., in preparation

\bibitem{Hwa04}
Hwang, U., Laming, J.M., Badenes, C., et al. 2004, ApJL, 615, 117

\bibitem[Kifonidis et al. 2003]{Kif03} Kifonidis, K., Plewa, T., Janka, H.-Th.,
\& M\"uller, E. 2003, A\&A 408, 621

\bibitem[Khokhlov et al. 1999]{Kh099} Khokhlov, A., H\"oflich, P., Oran, E., 
Wheeler, J. C., Wang, L. \& Chtchelkanova, A. Yu 1999, ApJL, 524, 107

\bibitem{Leo01}
Leonard, D. C., \& Filippenko, A. V. 2001, PASP, 113, 920

\bibitem{Nag00}
Nagataki, S. 2000, ApJS, 127, 141

\bibitem{Sch92}
Scheck, L., Plewa, T., Janka, H.-Th., Kifonidis, K., M\"uller, E., 2004,
PhRvL, 1992, 011103

\bibitem{Spy90}
Spyromilio, J., Meikle, W. P. S., \& Allen, D. A. 1990, 
MNRAS, 242, 669

\bibitem{Sun87}
Sunyaev, R., Kaniovskii, A., Efremov, V., Gilfanov, M., \& Churazov, E. 
1987, Nature, 330, 227

\bibitem{Tim00}
Timmes, F., Hoffman, R. \& Woosley, S., 2000, ApJS, 129, 377

\bibitem{Tul90}
Tueller, J., Barthelmy, S., Gehrels, N., Teegarden, B. J., 
Leventhal, M., \& MacCallum, C. J. 1990, ApJ, 351, L41

\bibitem{Wan01}
Wang, L., Howell, D. A., H\"oflich, P., Wheeler, J. C. 2001, 
ApJ, 556, 302

\bibitem{Wea93}
Weaver, T. A., \& Woosley, S. E. 1993, Phys. Rep., 227, 65

\bibitem{Wit89}
Witteborn, F.C., Bregman, J.D., Wooden, D.H., Pinto, P.A.,
Rank, D.M., Woosley, S.E., \& Cohen, M. 1989, ApJL, 338, L9

\bibitem{Woo88}
Woosley, S. E. 1988, ApJ, 330, 218


\end{thebibliography}
\end{document}